\documentclass[iop]{emulateapj}
\usepackage{enumerate}
\usepackage{graphicx}
\usepackage{amssymb}
\usepackage{epstopdf}
\usepackage{amsmath}
\usepackage[breaklinks]{hyperref}
\usepackage{ulem}
\newcommand{\imsize}{3in} 

\newcommand{\nRV}{26}

\newcommand{\starname}{KOI-94}
\newcommand{\planetb}{KOI-94b}
\newcommand{\planetc}{KOI-94c}
\newcommand{\planetd}{KOI-94d}
\newcommand{\planete}{KOI-94e}
\newcommand{\kicid}{KIC~6462863}
\newcommand{\tmid}{2MASS J19491993+4153280}

\newcommand{\eke}{\textit{Kepler}\ }

\newcommand{\kms}{\ensuremath{\rm km\,s^{-1}}}
\newcommand{\ms}{\ensuremath{\rm m\,s^{-1}}}

\newcommand{\gcmc}{\ensuremath{\rm g\,cm^{-3}}}
\newcommand{\gcc}{\gcmc}
\newcommand{\fluxunit}{\ensuremath{\rm erg\,s^{-1}\,cm^{-2}}}

\newcommand{\teff}{\ensuremath{T_{\rm eff}}}
\newcommand{\logg}{\ensuremath{\log{g}}}
\newcommand{\vsini}{\ensuremath{v \sin{i}}}
\newcommand{\feh}{[Fe/H]}

\newcommand{\rsun}{\ensuremath{R_\sun}}
\newcommand{\msun}{\ensuremath{M_\sun}}
\newcommand{\lsun}{\ensuremath{L_\sun}}

\newcommand{\rstar}{\ensuremath{R_\star}}
\newcommand{\mstar}{\ensuremath{M_\star}}
\newcommand{\loggstar}{\ensuremath{\logg_\star}}
\newcommand{\lstar}{\ensuremath{L_\star}}

\newcommand{\rhostar}{\ensuremath{\rho_\star}}

\newcommand{\rpl}{\ensuremath{R_{\rm P}}}
\newcommand{\mpl}{\ensuremath{M_{\rm P}}}

\newcommand{\rhopl}{\ensuremath{\rho_{\rm P}}}

\newcommand{\teq}{\ensuremath{T_{\rm eq}}}

\newcommand{\rjup}{\ensuremath{R_{\rm J}}}
\newcommand{\mjup}{\ensuremath{M_{\rm J}}}

\newcommand{\rearth}{\ensuremath{R_{\earth}}}
\newcommand{\mearth}{\ensuremath{M_{\earth}}}

\newcommand{\mspecial}{150 \mearth}

\newcommand{\teffSME}{\ensuremath{6182 \pm 58}}    
\newcommand{\fehSME}{\ensuremath{+0.0228 \pm 0.0020}}   
\newcommand{\loggSME}{\ensuremath{4.181 \pm 0.066}}   
\newcommand{\vsiniSME}{\ensuremath{7.3 \pm 0.5}}    
\newcommand{\mstarRowe}{\ensuremath{1.277 \pm 0.050}}
\newcommand{\rstarRowe}{\ensuremath{1.52 \pm 0.14}}

\newcommand{\lumstarRowe}{\ensuremath{3.01 \pm 0.60}}
\newcommand{\ageRowe}{\ensuremath{3.16 \pm 0.39}}
\newcommand{\KepMag}{\ensuremath{12.2}}

\newcommand{\periodb}{\ensuremath{3.743208 \pm 0.000015}}
\newcommand{\periodbshort}{3.74}
\newcommand{\epochb}{\ensuremath{2454964.6175 \pm 0.0021}}
\newcommand{\impactb}{\ensuremath{0.088 \pm 0.072}}
\newcommand{\scaledSemiMajb}{\ensuremath{7.25 \pm 0.59}}
\newcommand{\scaledPlanetRadiusb}{\ensuremath{0.01031 \pm 0.00014}}
\newcommand{\inclinationb}{\ensuremath{89.30 \pm 0.57}}

\newcommand{\periodc}{\ensuremath{10.423648 \pm 0.000016}}
\newcommand{\periodcshort}{10.4}
\newcommand{\epochc}{\ensuremath{2454971.00870 \pm 0.00103}}
\newcommand{\impactc}{\ensuremath{0.41 \pm 0.018}}
\newcommand{\scaledSemiMajc}{\ensuremath{14.3 \pm 1.2}}
\newcommand{\scaledPlanetRadiusc}{\ensuremath{0.02599 \pm 0.00047}}
\newcommand{\inclinationc}{\ensuremath{88.36 \pm 0.75}}

\newcommand{\periodd}{\ensuremath{22.3429890 \pm 0.0000067}}
\newcommand{\perioddshort}{22.3}
\newcommand{\epochd}{\ensuremath{2454965.74052 \pm 0.00015}}
\newcommand{\impactd}{\ensuremath{0.055 \pm 0.051}}
\newcommand{\scaledSemiMajd}{\ensuremath{23.8 \pm 1.9}}
\newcommand{\scaledPlanetRadiusd}{\ensuremath{0.068016 \pm 0.000080}}
\newcommand{\inclinationd}{\ensuremath{89.871     \pm    0.123}}

\newcommand{\periode}{\ensuremath{54.32031  \pm    0.00012}}
\newcommand{\periodeshort}{54.3}
\newcommand{\epoche}{\ensuremath{2454994.2379    \pm   0.0012}}
\newcommand{\impacte}{\ensuremath{0.18    \pm    0.11}}
\newcommand{\scaledSemiMaje}{\ensuremath{43.1      \pm    3.5}}
\newcommand{\scaledPlanetRadiuse}{\ensuremath{0.03960    \pm  0.00024}}
\newcommand{\inclinatione}{\ensuremath{89.76     \pm   0.15}}

\newcommand{\semiAmpb}{\ensuremath{3.3 \pm 1.4}}
\newcommand{\eccb}{\ensuremath{0.25 \pm 0.17}}
\newcommand{\gammaVelb}{\ensuremath{2.1 \pm 1.4}}
\newcommand{\trend}{\ensuremath{-0.0125 \pm 0.0063}}

\newcommand{\semiAmpc}{\ensuremath{3.5^{+1.3}_{-3.5}}}

\newcommand{\eccc}{\ensuremath{0.43 \pm 0.23}}

\newcommand{\semiAmpd}{\ensuremath{18.3 \pm 1.5}}
\newcommand{\eccd}{\ensuremath{0.022 \pm 0.038}}

\newcommand{\semiAmpe}{\ensuremath{4.5^{+2.3}_{-3.5}}}
\newcommand{\ecce}{\ensuremath{0.019 \pm 0.23}}

\newcommand{\mplanetb}{\ensuremath{10.5     \pm     4.6}}
\newcommand{\rplanetb}{\ensuremath{1.71     \pm    0.16}}
\newcommand{\rhoplanetb}{\ensuremath{10.1     \pm     5.5}}
\newcommand{\semiMajb	}{\ensuremath{0.05119   \pm   0.00067}}	
\newcommand{\teqb}{\ensuremath{1486}}	

\newcommand{\mplanetc}{\ensuremath{15.6^{+5.7}_{-15.6}}}
\newcommand{\rplanetc}{\ensuremath{4.32    \pm     0.41}}
\newcommand{\rhoplanetc}{\ensuremath{0.91^{+0.36}_{-0.91}}}
\newcommand{\semiMajc}{\ensuremath{0.1013  \pm     0.0013}}	
\newcommand{\teqc}{\ensuremath{1012}} 

\newcommand{\mplanetd}{\ensuremath{106    \pm      11}}

\newcommand{\rplanetd}{\ensuremath{11.27 \pm 1.06}}

\newcommand{\rhoplanetd}{\ensuremath{0.363 \pm 0.101}}
\newcommand{\semiMajd	}{\ensuremath{0.1684  \pm     0.0022}}	

\newcommand{\teqd}{\ensuremath{806}}	

\newcommand{\mplanete}{\ensuremath{35^{+18}_{-28}}}
\newcommand{\rplanete}{\ensuremath{6.56 \pm 0.62}}
\newcommand{\rhoplanete}{\ensuremath{0.60^{+0.26}_{-0.56}}}
\newcommand{\semiMaje}{\ensuremath{0.3046  \pm     0.0040}}	
\newcommand{\teqe}{\ensuremath{584}}	

\newcommand{\gammacirc}{\ensuremath{1.76 \pm 1.4}}
\newcommand{\semiAmpbcirc}{\ensuremath{3.2 \pm 1.7}}
\newcommand{\mplanetbcirc}{\ensuremath{9.2 \pm 4.9}}
\newcommand{\rhoplanetbcirc}{\ensuremath{9.0 \pm 4.7}}

\newcommand{\semiAmpccirc}{\ensuremath{1.6 \pm 1.3}}
\newcommand{\mplanetccirc}{\ensuremath{6.5 \pm 6.3}}
\newcommand{\rhoplanetccirc}{\ensuremath{0.46 \pm 0.37}}

\newcommand{\semiAmpdcirc}{\ensuremath{19.68 \pm 2.19}}
\newcommand{\mplanetdcirc}{\ensuremath{102 \pm 11.4}}
\newcommand{\rhoplanetdcirc}{\ensuremath{0.380 \pm 0.042}}

\newcommand{\semiAmpecirc}{\ensuremath{5.25 \pm 2.04}}
\newcommand{\mplanetecirc}{\ensuremath{36.6 \pm 14.4}}
\newcommand{\rhoplanetecirc}{\ensuremath{0.690 \pm 0.268}}

\newcommand{\rcoeffs}{\ensuremath{1.78~}}
\newcommand{\mcoeffs}{\ensuremath{0.53~}}
\newcommand{\fcoeffs}{\ensuremath{-0.03~}}

\newcommand{\rcoeffg}{\ensuremath{2.45~}}
\newcommand{\mcoeffg}{\ensuremath{-0.039~}}
\newcommand{\fcoeffg}{\ensuremath{0.094~}}

\newcommand{\dcoeffs}{\ensuremath{1.30~}}
\newcommand{\mdcoeffs}{\ensuremath{-0.60~}}
\newcommand{\fdcoeffs}{\ensuremath{0.09~}}

\newcommand{\dcoeffg}{\ensuremath{0.48~}}
\newcommand{\mdcoeffg}{\ensuremath{1.10~}}
\newcommand{\fdcoeffg}{\ensuremath{-0.28~}}

\newcommand{\rsolid}{\ensuremath{\frac{\rpl}{\rearth} = 
\rcoeffs
\left(\frac{\mpl}{\mearth}\right)^{\mcoeffs}
\left(\frac{F}{\fluxunit}\right)^{\fcoeffs}}}

\newcommand{\rgiant}{\ensuremath{\frac{\rpl}{\rearth} = 
\rcoeffg
\left(\frac{\mpl}{\mearth}\right)^{\mcoeffg}
\left(\frac{F}{\fluxunit}\right)^{\fcoeffg}}}

\newcommand{\forgiant}{\ensuremath{\text{for } \mpl > \mspecial}}
\newcommand{\forsolid}{\ensuremath{\text{for } \mpl < \mspecial}}

\newcommand{\rmsrg}{1.15}
\newcommand{\rmsrhog}{1.48}
\newcommand{\rmsrs}{1.41}
\newcommand{\rmsrhos}{2.69}

\newcommand{\medianflux}{\ensuremath{8.6 \times 10^{8}}}

\begin{document}

\title{The Mass of KOI-94\MakeLowercase{d} and a Relation for Planet Radius, Mass, and Incident Flux $\dagger$}
\author{
Lauren~M.~Weiss$^{1,\ddagger,\star}$,
Geoffrey~W.~Marcy$^1$, 
Jason~F.~Rowe$^2$,
Andrew~W.~Howard$^{3}$,
Howard~Isaacson$^1$, 
Jonathan~J.~Fortney$^4$, 
Neil~Miller$^4$, 
Brice-Olivier~Demory$^5$, 
Debra~A.~Fischer$^6$, 
Elisabeth~R.~Adams$^7$, 
Andrea~K.~Dupree$^7$,
Steve~B.~Howell$^2$, 
Rea Kolbl$^1$,
John~Asher~Johnson$^8$,
Elliott~P.~Horch$^9$,
Mark~E.~Everett$^{10}$,
Daniel~C.~Fabrycky$^{11}$,
Sara Seager$^5$
}

\affil{$^1$B-20 Hearst Field Annex, Astronomy Department, University of California, Berkeley, Berkeley, CA 94720}
\affil{$^2$NASA Ames Research Center, Moffett Field, CA 94035}
\affil{$^3$Institute for Astronomy, University of Hawaii, 2680 Woodlawn Drive, Honolulu, HI 96822, USA}
\affil{$^4$University of California, Santa Cruz, Department of Astronomy \& Astrophysics, 1156 High Street, 275 Interdisciplinary Sciences Building (ISB), Santa Cruz, CA 95064}
\affil{$^5$Massachusetts Institute of Technology, 77 Massachusetts Avenue, Cambridge, MA, 02139}
\affil{$^6$Department of Astronomy, Yale University, P.O. Box 208101, New Haven, CT 06510-8101}
\affil{$^7$Harvard-Smithsonian Center for Astrophysics, 60 Garden Street, Cambridge, MA 02138}
\affil{$^8$California Institute of Technology, 1216 E. California Blvd., Pasadena, CA 91106}
\affil{$^9$Southern Connecticut State University, Dept of Physics, 501 Crescent St., New Haven, CT 06515}
\affil{$^{10}$National Optical Astronomy Observatory, 950 N. Cherry Ave, Tucson, AZ 85719}
\affil{$^{11}$The Department of Astronomy and Astrophysics, University of Chicago, 5640 S. Ellis Ave, Chicago, Illinois 60637}
\altaffiltext{$\star$}{\small To whom correspondence should be addressed.  E-mail: lweiss@berkeley.edu}
\altaffiltext{$\dagger$}{\small Based in part on observations obtained at the W.~M.~Keck Observatory, which is operated by the University of California and the California Institute of Technology.}
\altaffiltext{$\ddagger$}{\small Supported by the NSF Graduate Student Fellowship, Grant DGE 1106400.}


\slugcomment{Draft: Submitted to ApJ 31 on Oct. 2012}


\begin{abstract}
\small We measure the mass of a modestly irradiated giant planet, \planetd. We wish to determine whether this planet, which is in a 22-day orbit and receives 2700 times as much incident flux as Jupiter, is as dense as Jupiter or rarefied like inflated hot Jupiters.  \starname\ also hosts at least 3 smaller transiting planets, all of which were detected by the \eke Mission.  With 26 radial velocities of \starname\ from the W.~M.~Keck Observatory and a simultaneous fit to the \eke light curve, we measure the mass of the giant planet and determine that it is not inflated.  Support for the planetary interpretation of the other three candidates comes from gravitational interactions through transit timing variations, the statistical robustness of multi-planet systems against false positives, and several lines of evidence that no other star resides within the photometric aperture.  We report the properties of \planetb\ (\mpl = \mplanetb \mearth, \rpl = \rplanetb \rearth, P = \periodbshort\ days), \planetc\ (\mpl = \mplanetc \mearth, \rpl = \rplanetc \rearth, P = \periodcshort\ days), \planetd\ (\mpl = \mplanetd \mearth, \rpl = \rplanetd \rearth P = \perioddshort\ days), and \planete\ (\mpl = \mplanete \mearth, \rpl = \rplanete \rearth, P = \periodeshort\ days).  The radial velocity analyses of \planetb\ and \planete\ offer marginal ($>2\sigma$) mass detections, whereas the observations of \planetc\ offer only an upper limit to its mass. Using the \starname\ system and other planets with published values for both mass and radius (138 exoplanets total, including 35 with $\mpl < \mspecial$), we establish two fundamental planes for exoplanets that relate their mass, incident flux, and radius from a few Earth masses up to ten Jupiter masses: $\rsolid \forsolid$, and $\rgiant \forgiant$.  These equations can be used to predict the radius or mass of a planet.
\end{abstract}




\section{Introduction}

One of the most pressing problems in exoplanetary physics is the anomalously large radii of close-in transiting gas giant planets.  These hot Jupiters have radii larger than predicted by standard models of giant planet cooling and contraction \citep[for reviews, see][]{Fortney10,Baraffe10}.  Some mechanism, or a variety of mechanisms, prevents planets from contracting, resulting in the observed inflated planetary radii.  The reasons for the radius anomaly for these planets could be tied to their formation and subsequent orbital evolution to close-in orbits, where the planets are subject to extremes of both tidal and radiative forcing.  Mechanisms to explain the large radii of the planets have tried to tap the vast energy sources available in tidal or radiative forcing.  It is critical to build up a large sample size of transiting giant planets, at a variety of orbital distances and incident fluxes, to better understand the physics that leads to the radius anomaly.

\citet{Miller11} pointed out that all transiting giant planets receiving less incident flux than $2\times 10^8$ erg s$^{-1}$ cm$^{-2}$ do not appear inflated, meaning that they are all smaller in radius than expected for pure H/He objects.  \citet{Miller11} estimated the masses of heavy elements contained within that relatively cool sample of 16 planets (at that time).  \citet{Demory11} extended this work to \eke gas giant candidate planets, and also found a lack of inflated candidates beyond this same critical flux level, $2\times 10^8$ erg s$^{-1}$ cm$^{-2}$.  This incident flux is approximately equal to an equilibrium temperature of 1000 K, for a zero Bond albedo and planet-wide redistribution of absorbed stellar flux.  The detailed study of giant planets receiving less than $2\times 10^8$ erg s$^{-1}$ cm$^{-2}$ in incident flux will serve as a useful contrast against the population of inflated giant planets that receive higher levels of incident flux.

To probe the underlying physical processes that cause the observed diversity of planetary densities, we need to both expand our sample and to test links between the planets' physical properties and their orbital properties. The first part of the paper focuses on expanding our sample. We measure the mass of a modestly irradiated giant planet or Òwarm Jupiter,Ó KOI-94d, in order to calculate its density and place constraints on its interior structure. We wish to determine whether this planet, which is in a \perioddshort-day orbit and receives 2675 times as much incident flux as Jupiter (just a bit below the ``critical" flux limit described above), is more similar to the bloated hot Jupiters or the cooler non-inflated gas giants, like our own Jupiter.

In addition to the warm Jupiter, KOI-94 hosts at least 3 smaller planets, all of which were detected through transit signatures in the photometry from the \eke Mission.  \citet{Hirano2012} note the planet-planet eclipse that occurs in this system, which allows a detailed analysis of the planets' orbital dynamics.  The multiplicity of this system presents an opportunity to examine the architecture of a closely packed system with a warm Jupiter.  Using Keck-HIRES radial velocities (RVs), we measure the mass of the warm Jupiter.  We obtain marginal mass measurements of two other planets in the system and an upper limit to the mass of the fourth.  Coupled with the mass of the giant planet we obtain from RVs, transit timing variations (TTVs) in the photometry allow an additional check of the RV masses obtained in this work.

In the second part of this paper, we investigate how the mass and incident flux of a planet relate to the planet's radius.  \citet{Enoch2012} and \citet{Kane2012} have done similar work, but here we include many more low-mass planets (down to 3 \mearth), allowing us to probe the mass-radius-flux relation at lower masses than in either of those papers.

This paper is structured as follows: \S2 reviews mechanisms for inflating giant planets, \S3 presents observations of \starname, \S4 describes the analyses used to derive planet masses, \S5 argues for the planetary status of all four transiting candidates of \starname\, \S6 describes the composition of \planetd, \S7  presents the radius-mass-incident flux relations for exoplanets and discusses possible interpretations, and \S8 summarizes the paper.



\section{Inflation Mechanisms}
A menagerie of radius inflation mechanisms have been proposed to explain the large radii of hot Jupiters.  The most recent reviews, now becoming slightly out of date, are in \citet{Fortney10} and \citet{Baraffe10}.  Without going into detail here on any one mechanism, we classify possible explanations into three groups: incident flux-driven mechanisms, tidal mechanisms, and delayed contraction.  

Some inflation mechanisms are driven by incident flux from the parent star, also called insolation, whereby a small fraction of the absorbed stellar flux is transported by a physical mechanism much deeper into the atmosphere, near or past the radiative-convective boundary.  These mechanisms are discussed in a variety of papers:  weather layer kinetic energy transport \citep{Showman02,Guillot02}, Ohmic dissipation \citep{Batygin10,Perna10}, thermal tides \citep{Arras10}, and mechanical greenhouse \citep{Youdin10}.  These mechanism would in general affect \emph{all} close in giant planets to some degree, with the strength of the effect waning at lower insolation levels.  Diversity in planetary mass, planetary heavy element masses, and planetary temperature would lead to a range of inflated radii.  For reference, the time-averaged incident flux on a planet is:
\begin{equation}
\langle F \rangle = \sigma \teff^4 \frac{\rstar^2}{4\pi a^{2}} \sqrt{\frac{1}{1-e^2}},
\label{eqn:incident_flux}
\end{equation}
where \rstar\ is the stellar radius, \teff\ is the effective stellar temperature, $a$ is the semi-major axis, and $e$ is the orbital eccentricity.

Another class of solutions are tidal interactions between the host star and planet, in particular, eccentricity damping.  Coupled tidal and planetary structure evolution has been calculated by a number of authors \citep{Bodenheimer2001,Gu03,Ibgui09,Miller09,Ibgui2010,Leconte10}.  The emerging view, in particular advanced by \citet{Leconte10} who used the most detailed tidal evolution equations, is that since radius inflation via orbital eccentricity damping is a transient phenomenon, it cannot be the ``universal'' radius inflation mechanism.  Radius inflation by tidal heating is a short-lived phenomenon, but the average system age is several Gyr.  However, in certain circumstances, including when a nonzero orbital eccentricity is maintained by outside forcing, tidal heating can inflate giant planet radii for as long as the forcing lasts.  The tidal power on a planet scales like:

\begin{equation}
\dot{E} = \frac{63}{4} \frac{(2\pi)^{5}}{G} \frac{\rpl^5}{Q'_P}\frac{e^2}{P^5}
\end{equation}
where $\dot{E}$ is the power, \rpl\ is the planetary radius, $G$ is Newton's gravitational constant, $Q'_P$ is the tidal dissipation factor, and $P$ is the orbital period \citep[rewritten from Equation 1 in][using Kepler's Third Law]{Ibgui2010}.  Two advantages of rewriting this equation are that (1) orbital period, rather than semi-major axis, is an observable, and (2) this formulation of the equation does not depend on stellar mass.

For completeness, we mention that delayed planetary contraction, due either to higher than anticipated atmospheric opacities \citep{Burrows07} or interior barrier to convection \citep{Chabrier07} are another class of solutions which could contribute somewhat to larger radii.  However, neither of these ideas should depend on proximity to the parent star, which should clearly be a feature of the correct solution(s).

\section{Observations}
In this section, we present photometry from the \eke Space Telescope, as well as data from the ground-based techniques of adaptive optics imaging, speckle imaging, and spectroscopy.  Transits of the planet candidates were identified in the \eke light curves in \citet{Batalha2012} and \citet{Borucki2012}.  In \S4, we describe our method for simultaneously fitting the photometry and radial velocities derived from time-series spectroscopy.  Here, we present adaptive optics imaging, speckle imaging, and spectroscopy of \starname.  These observations rule out various instances of a nearby, stellar companion that could masquerade as a transiting planet, or possible false-positive scenarios, as discussed in \S5.

\subsection{\eke Space Telescope}
\begin{figure}[htbp] 
   \centering
   \includegraphics[width=\imsize]{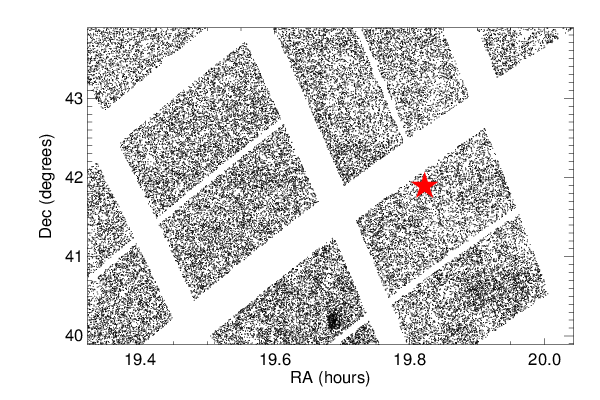} 
   \caption{\footnotesize A section of the \eke footprint.  Each point is a \eke target.  The red five-pointed star marks the location of KOI-94.}
   \label{fig:footprint}
\end{figure}

KOI-94 is on the edge of one of the chips of the \eke CCD (see Figure \ref{fig:footprint}), so when the spacecraft rotates, the CCD loses the part of its field of view containing \starname.  This causes two-quarter long gaps in the light curve.

\begin{figure*}[htbp] 
\centering
   \includegraphics[angle=270,width=6in]{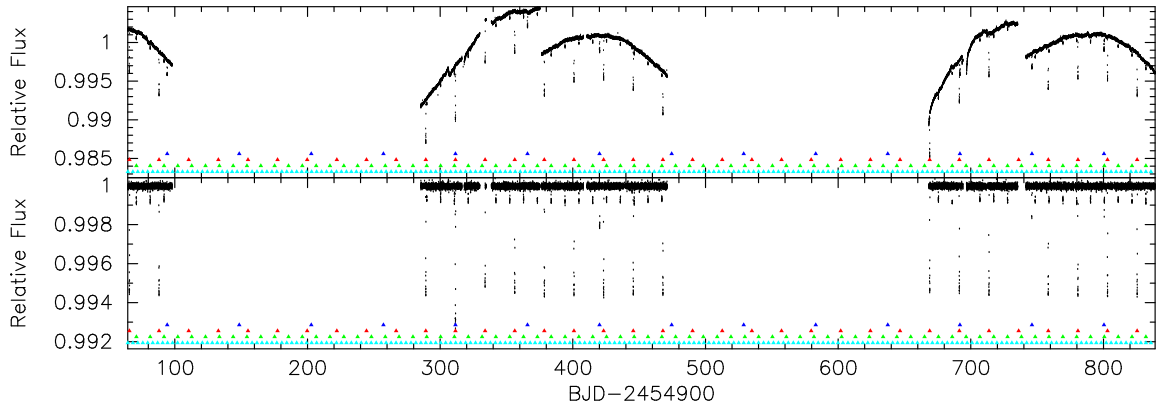} 
   \caption{\footnotesize The top panel shows the aperture photometry long cadence (30 minute) light-curve.  No corrections have been applied to the photometric measurements.  The bottom panel shows the detrended light-curve.  The data were detrended by applying a 2-day median filter.  Observations that occurred during a planetary transit were excluded from the calculation of the median.  The triangles indicate when transits occurred for planets b, c, d, and e with corresponding colours of cyan, green, red and blue.}
   \label{fig:jasonrowe_lc}
\end{figure*}

\begin{figure}[htbp] 
   \centering
   \includegraphics{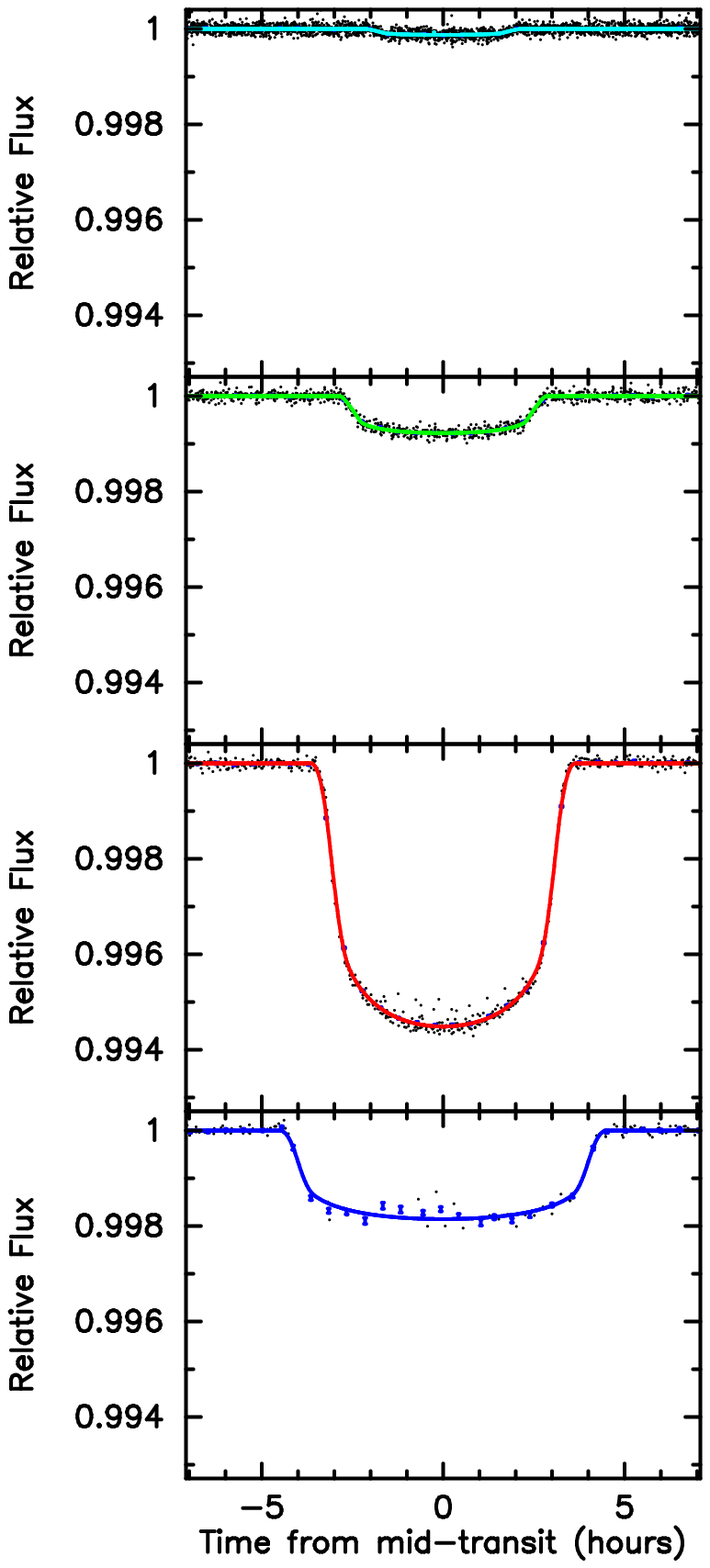} 
   \caption{\footnotesize The top panel shows the photometric observations phased to the orbital period of planet b with the best fit transit model overlaid.  The transit models for planets c, d and e have been removed. The transit times have been corrected for transit-timing variations. The next three panels show the transit lightcurves centred on planets c, d, and e respectively.  As in the upper panel, the best fit models for other transiting planet have been removed and corrected for transit timing variations.  The colors of the overlaid models match the identification used in Figure \ref{fig:jasonrowe_lc}.}
   \label{fig:jasonrowe_ph}
\end{figure}

Nonetheless, the \eke pipeline, as described in \citet{Batalha2012}, has found 4 transiting planet candidates associated with KOI-94 (see Figure \ref{fig:jasonrowe_lc}).  The light curve is phase-folded around their transit centers in Figure \ref{fig:jasonrowe_ph}.  In summary, they are a super-Earth (\planetb) in a 3.7-day orbit, a mini-Neptune or super-Earth (\planetc) in a 10-day orbit, a Jupiter-size planet (\planetd) in a 22-day orbit, and a Neptune-size planet (\planete) in a 54-day orbit.  The Kepler Input Catalog (KIC) ascribes an effective temperature of 6217 K and a radius of 1.238 \rsun to \starname, resulting in planetary radii of 1.41 \rearth, 3.44 \rearth, 9.26 \rearth, and 5.48 \rearth for planets b, c, d, and e, respectively.  Our analysis of the stellar spectrum, which finds different values for the stellar temperature and radius and therefore for the planet radii, is described in \S4.

\subsection{Adaptive Optics}
\begin{figure*}[htbp] 
   \centering
   \includegraphics[width=6in]{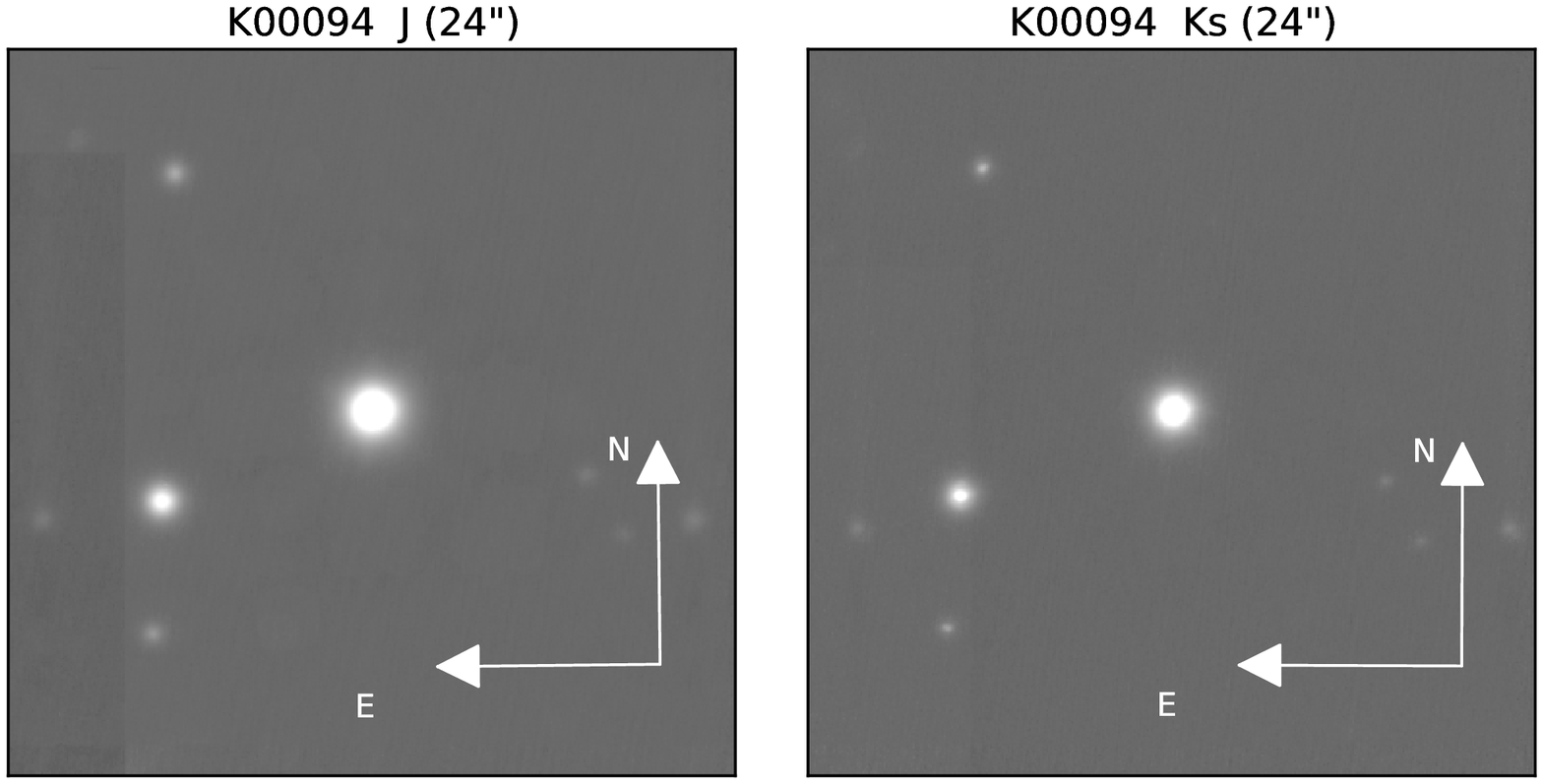} 
   \caption{\small Adaptive optics image of KOI-94 (center) in J and Ks. The
closest companion is 7.5\arcsec\ away.  All of the objects in the frame are stars except for the spot in the bottom left, which is an artifact.
}
   \label{fig:ao}
\end{figure*}

KOI-94 was observed with near-infrared adaptive optics on 2009 Nov 08
using ARIES on the MMT \citep{Adams2012}. Images were obtained in both
J and Ks, and reveal no companions closer than 7.5\arcsec\ ($\Delta_J
= 2.5$, $\Delta_{Ks} = 2$; see Figure \ref{fig:ao}).  The image FWHM was 0.23 in Ks and 0.43 in
J. We can place a limit on undetected companions of $\Delta_{J} =2.2$
and $\Delta_{Ks} =3.4$ at 0.5\arcsec,  $\Delta_{J} =4.6$ and
$\Delta_{Ks} =5.9$ at 1.0\arcsec, and  $\Delta_{J} =8.7$ and
$\Delta_{Ks} =9.1$ at 4.0\arcsec (and beyond). Any additional
companions that would dilute the transit light curve and change the
planet parameters are constrained to be faint or very close to the
star.

\subsection{Speckle Imaging}
\begin{figure}[htbp] 
   \centering
   \includegraphics[trim=35mm 20mm 35mm 0mm, clip,width=\imsize]{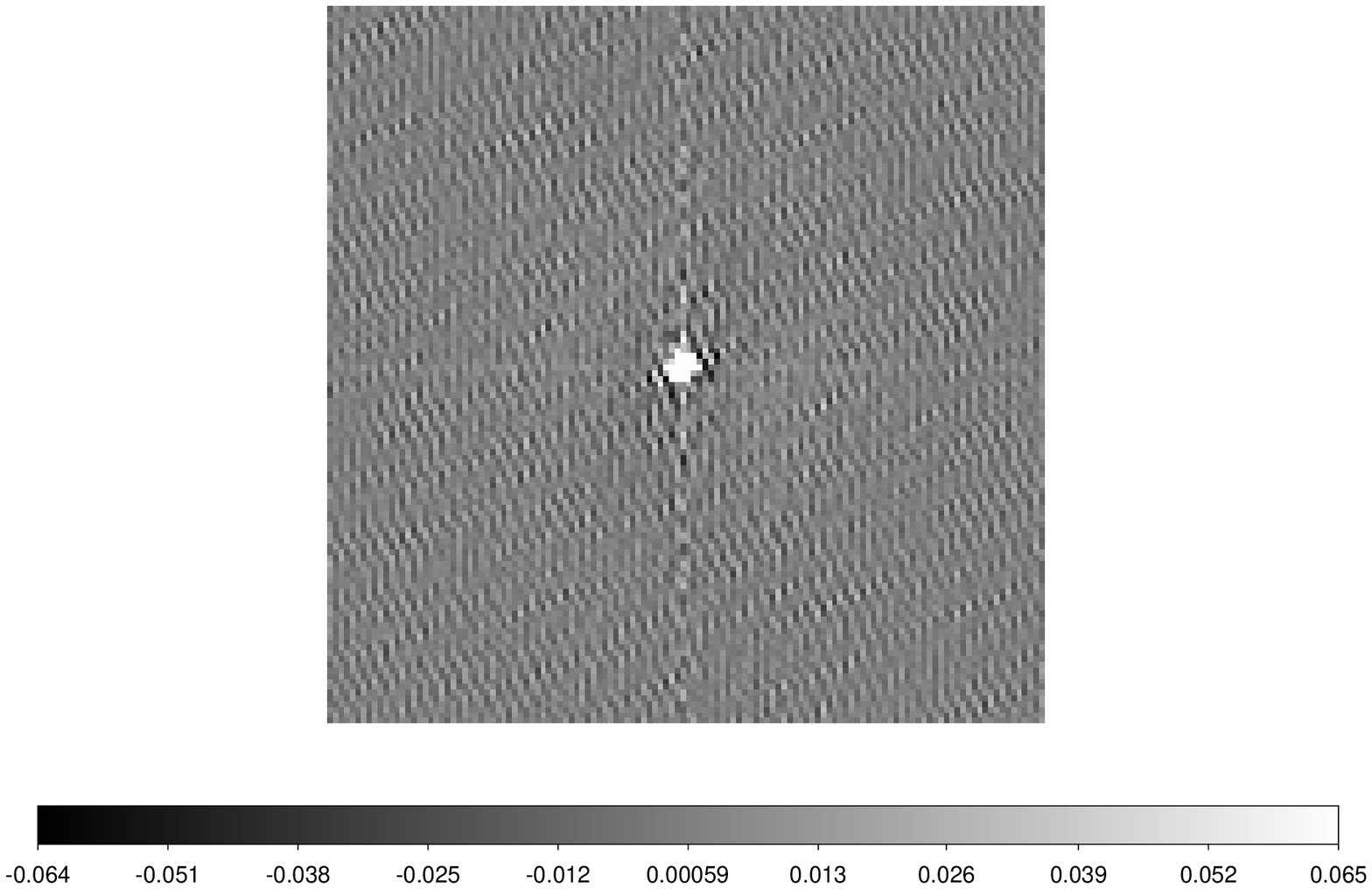} 
   \caption{\footnotesize Speckle image obtained 2012-10-23 at the WIYN telescope at 692 nm.  The image spans 2.76\arcsec $\times$ 2.76\arcsec.}
   \label{fig:speckle}
\end{figure}

Speckle imaging of KOI-94 was obtained on the night of 19 June 2010 UT and the night of 23 Oct 2010 using the two-color DSSI speckle camera at the WIYN 3.5-m telescope on Kitt Peak. The speckle camera simultaneously obtained 5000 (3000) 40 msec images on 19 June (23 Oct) in filters: $V$ (center = 5620\AA, width = 400\AA) , $R$ (center = 6920\AA, width = 400\AA) and $I$ (center = 8880\AA, width = 400\AA). These data were reduced and processed to produce a final reconstructed speckle image for each filter. Figure~\ref{fig:speckle}  shows the reconstructed R band image.  North is up and East is to the left in the image and the ``cross'' pattern seen in the image is an artifact of the reconstruction process.  The details of the two-color speckle camera observations and the \eke follow-up observing program are presented in \citet{Howell2011}.

On both occasions, the speckle data for this $R$=12.5 star allow detection of a companion star within the approximately $2.76 \times 2.76$\arcsec\ box centered on the target.  The speckle observation can detect, or rule out, companions between 0.05\arcsec\ and 1.5\arcsec\ from \starname.  The June 2010 speckle image was obtained with the WIYN telescope during relatively poor native seeing near 1.0\arcsec, while the Oct 2010 observations made during good seeing, 0.6\arcsec. We found no companion star within the speckle image separation detection limits to a delta magnitude limit of $\sim$4 mag in the R band, 2 mag in the V band, and 3.4 mag in the I band.

\subsection{Spectroscopy}
We obtained time-series spectroscopy of \starname\ on the W.~M.~Keck I telescope with the HIRES echelle spectrometer through an iodine cell.  Most of the observations occurred between May and August 2012, but our earliest observation was 2009-12-07.  We rejected spectra with fewer than 4000 ADU (8760 photons), since these low-signal spectra resulted in large ($> 6$ m/s) radial velocity errors.  We also excluded spectra taken during a transit of \planetd\ to avoid confusion from the Rossiter-McLaughlin effect.  This yielded \nRV\ spectra that we deemed suitable for RV analysis, as presented in Table \ref{tab:obstable}.  We also obtained a spectrum of \starname\ without the iodine cell as a template for RV analysis and for characterizing the star (see \S4).

 Note that one observation, on JD  16135.095, occurred during a transit of planet e.  To see whether this datum affected our mass results, we tried removing it and repeating the circular analysis described in \S4.2.  Removing the datum changed the masses of the planets by $\sim.01\sigma$, and so we can safely ignore the effects of this datum on our analysis of the planetary system.

\begin{deluxetable}{rrrr}
\tabletypesize{\scriptsize}
\tablewidth{0pt}
\tablecaption{Spectroscopic Observations \label{tab:obstable}}
\tablehead{\colhead{JD - 2440000.0} & \colhead{Photons} & \colhead{RV} & \colhead{Error} \\ 
\colhead{} & \colhead{} & \colhead{(m/s)} & \colhead{(m/s; Jitter = 3.0)} } 

\startdata
 15172.768 &       9351 &       -9.6 &        6.1 \\ 
 16076.070 &      11709 &       27.0 &        5.5 \\ 
 16100.047 &      23095 &       14.2 &        5.0 \\ 
 16109.929 &      23065 &      -13.5 &        5.0 \\ 
 16111.048 &      22920 &      -16.8 &        5.0 \\ 
 16113.043 &      22594 &      -14.9 &        5.0 \\ 
 16114.077 &      14484 &      -17.0 &        6.0 \\ 
 16115.074 &      21584 &       -5.0 &        5.0 \\ 
 16116.054 &      23041 &        8.1 &        4.8 \\ 
 16134.015 &      22973 &       -9.2 &        4.9 \\ 
 16135.095 &      23592 &       -9.7 &        4.7 \\ 
 16139.015 &      22594 &        6.8 &        4.5 \\ 
 16140.813 &      22370 &        3.8 &        5.0 \\ 
 16144.100 &      20178 &       20.3 &        5.2 \\ 
 16144.796 &      23330 &       14.6 &        4.6 \\ 
 16146.057 &      18774 &       18.5 &        4.9 \\ 
 16147.972 &      22642 &        2.5 &        4.7 \\ 
 16149.091 &      15395 &      -13.8 &        5.0 \\ 
 16149.752 &      22918 &        6.1 &        4.5 \\ 
 16149.766 &      22629 &       -6.7 &        4.6 \\ 
 16150.065 &      19204 &       -2.6 &        4.8 \\ 
 16150.079 &      19162 &       -5.6 &        4.8 \\ 
 16150.094 &      18323 &      -10.1 &        4.9 \\ 
 16151.075 &      22791 &      -10.0 &        5.2 \\ 
 16164.001 &      23183 &       23.0 &        4.9 \\ 
 16173.925 &      21904 &      -10.7 &        5.1 \\ 
\enddata


\tablecomments{Observations with low signal-to-noise  (fewer than 8760 photons) were not used in our analysis and are omitted from this table.  Photon counts are averaged over all pixels.}


\end{deluxetable}

We determined the radial velocity of \starname\ in each iodine spectrum based on the spectrum's Doppler shift.  We used the lab-frame iodine lines that the iodine cell superimposes on the stellar spectrum to calibrate velocities to an instrumental precision of 3~\ms (although stellar jitter introduces additional errors).  This technique is described in further detail in \citet{Howard2011}.

\section{Planet and Stellar Properties from Radial Velocities, Photometry and Spectra}
\begin{figure}[htbp] 
   \centering
   \includegraphics[width=\imsize]{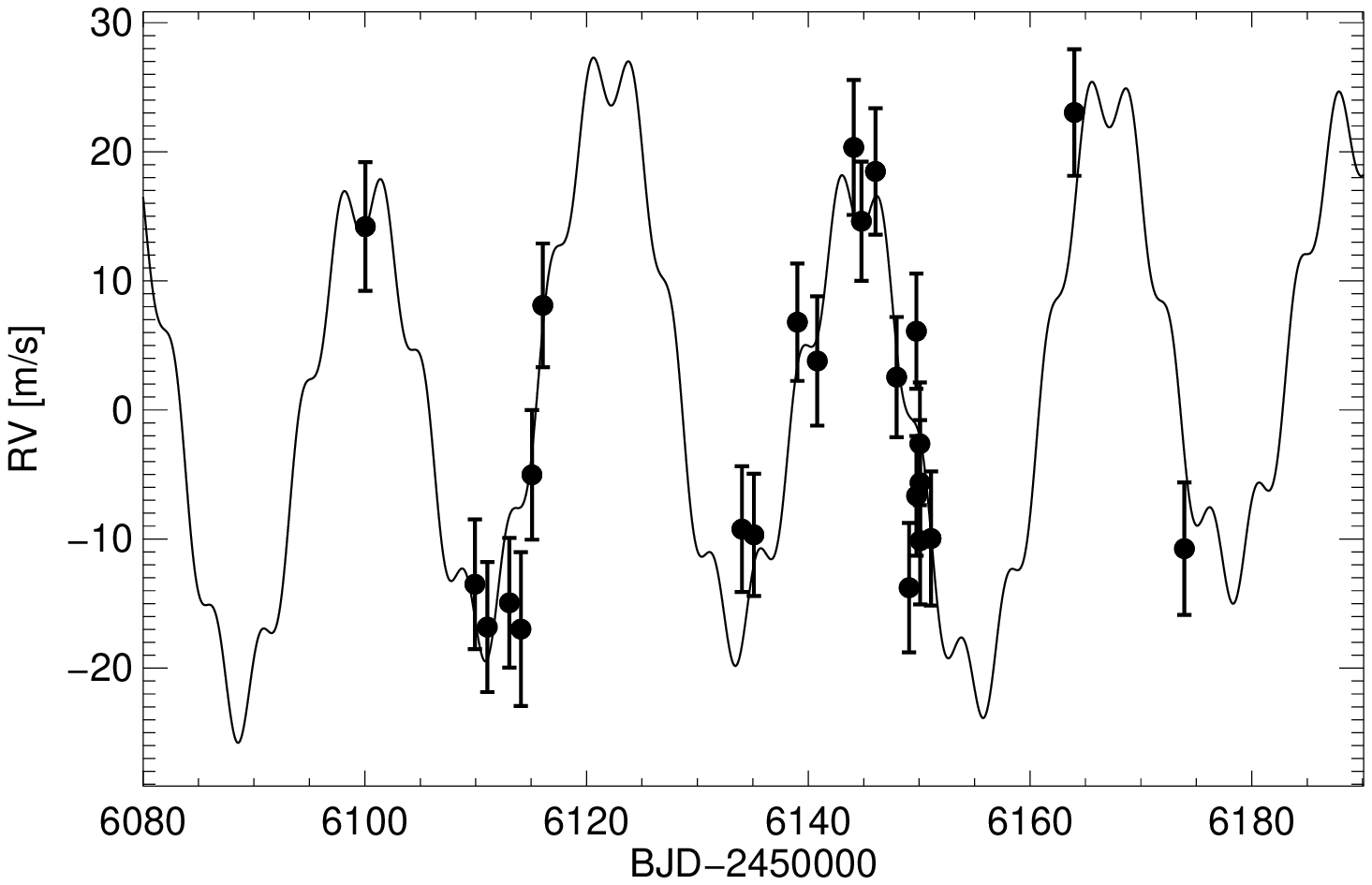}
   \caption{\footnotesize Radial velocity versus time from May 2012 onward.  Black points are data with 1$\sigma$ errors (assuming a stellar jitter of 3.0 m/s); a circular four-planet fit is superimposed.}
   \label{fig:rvcurve}
\end{figure}

We use radial velocity measurements of KOI-94 to determine the mass of planet d, marginal mass detections of planets e and b, and an upper limit on the mass of planet c.  A plot of radial velocity versus time and a four-planet circular fit is shown in Figure \ref{fig:rvcurve}.  

We fit the RVs with three models, each of which provides an interesting interpretation of the system.  The first model has only one planet (the giant) in a circular orbit.  This is because the giant planet dominates the RVs, and so it is useful to compare a four-planet solution to a simpler one-planet solution.   In our second model, all four planets are in circular orbits.  Because KOI-94 is a closely-packed system, we do not expect large eccentricities of the planets, so we want to verify that a solution allowing eccentricities is not significantly different from a circular solution.  In our third and most sophisticated model, we fit for all four planets in eccentric orbits while simultaneously fitting the light curve.  

\subsection{Circular Orbit Solutions}
\begin{figure}[htbp] 
   \centering
   \includegraphics[width=\imsize]{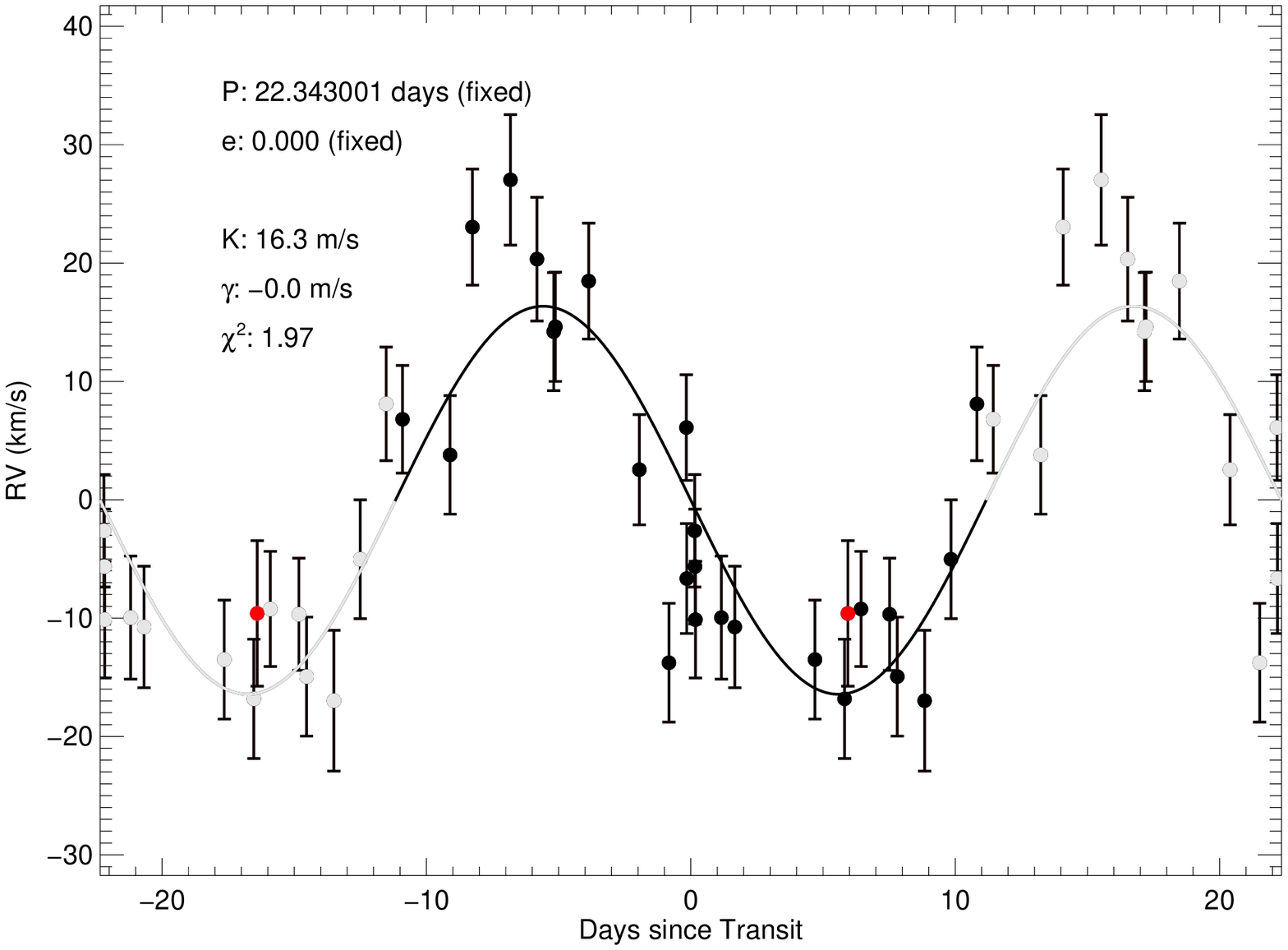}
   \caption{\footnotesize A one-planet, circular fit to the RVs, phase-folded to the period of KOI-94d.  The black points are the data (error bars are $1\sigma$), and the black line is the circular one-planet fit to the data.  The gray points and fit are time-shifted repetitions of the black data points and fit.  The red point is the oldest data point (2009); all other data are from summer 2012.}
   \label{fig:rv_oneplanet}
\end{figure}

Here we compare the results of the one-planet and four-planet circular orbital solutions.  For the one-planet fit, we find $K = 16.3\,\ms$, producing a reduced $\chi^2$ of 1.97.  The data and phase-folded, one-planet circular fit are shown in Figure \ref{fig:rv_oneplanet}.  

\begin{figure*}[htbp] 
\centering
   \includegraphics[width=6in]{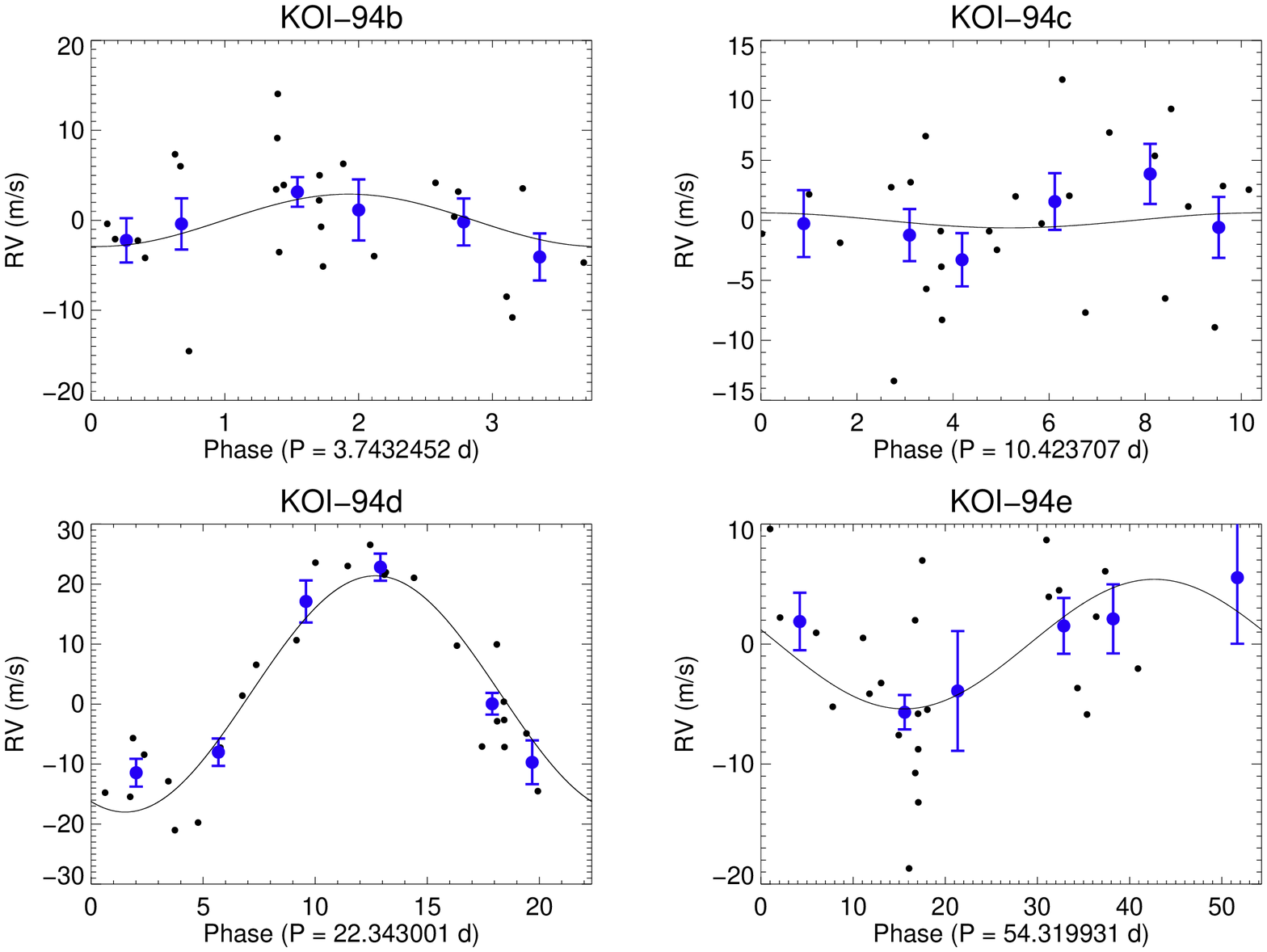}
   \caption{\footnotesize Radial velocity components from a four-planet circular fit.  Each panel shows the radial velocity signature from one planet.  The black line is the model fit; the black points are the RV data minus the model RVs from the other three planets.  The blue points are the binned RVs.  We chose the number of bins by rounding up the square root of the number of observations, creating 6 bins of equal spacing in phase.  The error bars are the uncertainty in the mean of the data in each bin.  The reduced $\chi^2$ of this fit is 1.60.  The RVs provide a $9\sigma$ detection of \planetd, a $2.5\sigma$ detection of \planete, a $2\sigma$ detection of \planetb\ and an upper limit on the mass of \planetc.}
   \label{fig:rv_phase}
\end{figure*}

In the four-planet circular model, the center-of-mass velocity $\gamma$ of the system and the semi-amplitude $K_n$ of each planet $n$ are allowed to vary, allowing 5 degrees of freedom; all other orbital parameters are fixed.  We augmented the photon-noise errors by 3.0 \ms in quadrature to account for the stellar jitter.  The best four-planet fit had a reduced $\chi^2$ of 1.60.  The best-fit radial velocity components from the four planets are shown in Figure \ref{fig:rv_phase}.  

Because the stellar jitter is unknown, we recalculated this fit, varying the stellar jitter to achieve $\chi^2=1$ (this yielded a jitter of 5.0 \ms).  The RV semi-amplitudes achieved in this fit were consistent with those assumed for a stellar jitter of 3.0 \ms: the semi-amplitude for \planetd\ and \planete\ changed by less than half a percent, and the semi-amplitudes for \planetb\ and \planetc\ fell within the $1\sigma$ errors.  We also note that because these are circular fits, the difference in $\chi^2$ between a model with a stellar jitter of 3.0\ms\ and 5.0\ms\ could arise from planetary eccentricities that are excluded from the model.  Therefore, we assume a stellar jitter of 3.0 \ms\ and adopt the resulting parameters for the 4-planet circular solution, which are reported in Table \ref{tab:SystemParamsCirc}.

\begin{deluxetable}{lc}
\tabletypesize{\scriptsize}
\tablewidth{0pc}
\tablecaption{Planet parameters for circular orbits of \starname. \label{tab:SystemParamsCirc}}
\tablehead{\colhead{Parameter} & \colhead{Value}} 
\startdata
\sidehead{\em Circular Keplerian Fit: \planetb}
Center-of-mass velocity $\gamma$ (\ms)           & \gammacirc \\
Orbital semi-amplitude $K$ (\ms)		& \semiAmpbcirc \\
Mass \mpl\ (\mearth)				& \mplanetbcirc \\
Density \rhopl\ (\gcmc)				& \rhoplanetbcirc \\

\sidehead{\em Circular Keplerian Fit: \planetc}
Orbital semi-amplitude $K$ (\ms)		& \semiAmpccirc \\
Mass \mpl\ (\mearth)				& \mplanetccirc \\
Density \rhopl\ (\gcmc)				& \rhoplanetccirc \\

\sidehead{\em Circular Keplerian Fit: \planetd}
Orbital semi-amplitude $K$ (\ms)		& \semiAmpdcirc \\
Mass \mpl\ (\mearth)				& \mplanetdcirc \\
Density \rhopl\ (\gcmc)				& \rhoplanetdcirc \\

\sidehead{\em Circular Keplerian Fit: \planete}
Orbital semi-amplitude $K$ (\ms)		& \semiAmpecirc \\
Mass \mpl\ (\mearth)				 & \mplanetecirc \\
Density \rhopl\ (\gcmc)				& \rhoplanetecirc \\
\enddata
\tablecomments{The best fit parameters for four planets in circular Keplerian orbits, after adopting stellar parameters and orbital ephemerides from the eccentric solution described in \S4.2 (see Table \ref{tab:SystemParams}).}
\end{deluxetable}

\begin{figure*}[htbp] 
   \centering
   \includegraphics[width=6in]{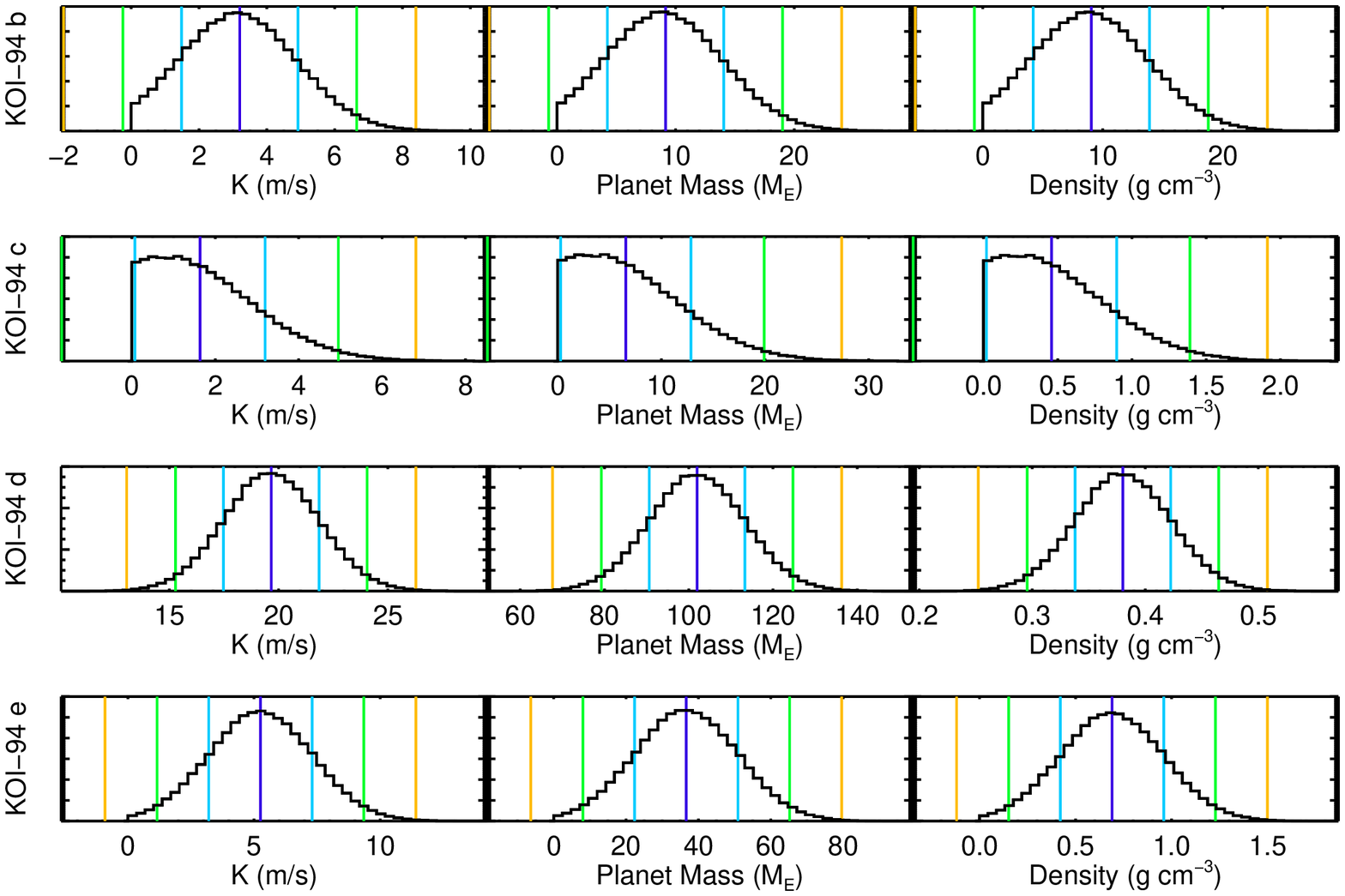} 
   \caption{\footnotesize Posterior distributions for the four-planet circular fit.  From left to right: likelihoods of $K$, $M_P$, and $\rho_P$ for (top to bottom) \planetb, \planetc, \planetd, and \planete.  The median, $1\sigma$, $2\sigma$, and $3\sigma$ likelihoods are over-plotted in dark blue, cyan, green, and orange.  To account for the positive definite bias of the likelihood distribution, $\sigma$ is measured above the median only; lower bounds are reflections of the upper bounds (i.e., a symmetric distribution is assumed).  This enables a quantitative estimate of the number of trials consistent with a non-detection via RVs.  Note that the best-fit values are obtained by minimizing $\chi^2$; the posterior distributions are used only for estimating uncertainties.  The best-fit values and uncertainties for the circular orbits are presented in Table \ref{tab:SystemParamsCirc}.}
   \label{fig:mcmc}
\end{figure*}

We assessed the errors in $K_n$ with a Markov Chain Monte Carlo (MCMC) analysis of 1 chain of 10$^7$ trials using the Metropolis-Hastings algorithm.  Figure \ref{fig:mcmc} shows the posterior likelihood distributions.  The corresponding planet mass (\mpl) and density (\rhopl) distributions (which are calculated from the orbital period and planet radius) are also shown.  These parameters are reported in Table \ref{tab:SystemParamsCirc}.

To calculate the probability of a non-detection in RVs for each planet, we assume a Gaussian posterior distribution and calculate the fraction of trials that would have occurred for masses at or below zero, had we not imposed a positive definite value for $K_n$.   We calculate the following probabilities of a non-detection via RVs: 0.04 for planet b, 0.14 for planet c, $5\times10^{-19}$ for planet d, and 0.007 for planet e.  However, these estimates of the non-detection probability are smaller than the true probability of non-detection because of the Lutz-Kelker bias, especially for planet c.

\subsubsection{Limiting Outer Planets}
The absence of a significant change in the radial velocity of the star between our earliest and most recent measurements strongly limits the possibility of a massive outer companion.  To quantify this, we computed a circular, four-planet orbit solution in which we allowed a linear trend in the radial velocity as a free parameter and used an MCMC analysis to explore the likelihood of a given linear velocity trend.  Over a two-year baseline, the median trend was \trend\, \ms\ per day, which corresponds to a $3\sigma$ upper limit of -6.9 \ms\ per year.  In the style of \citet{Winn2010}, we compute the mass of an outer perturber based on the stellar acceleration, $\dot{\gamma}$, assuming the planet induces Newtonian gravitational acceleration on the star and in the limit $\mpl \ll \mstar$: $\dot{\gamma} = G\mpl/a^2$. To induce a stellar acceleration $\dot{\gamma}$ of -6.9 \ms\ per year via Newtonian gravity, an outer perturber would need to satisfy
\begin{equation}
\frac{\mpl \mathrm{sin}i}{\mjup} \left(\frac{a}{10\ \mathrm{AU}}\right)^{-2} = 3.9,
\label{eqn:perturber}
\end{equation}
where $i$ is the inclination of the planet's orbital plane with respect to the line of sight and \mjup\ is the mass of Jupiter.  Thus, with a significance of $3\sigma$, we can rule out companions more massive than 3.9 \mjup\ within 10 AU or more massive than 1.0 \mjup\ within 5 AU.

\subsection{Eccentric Orbit Solution}
The four-planet fit in which we allow eccentricities to float is the most versatile model.  This model has the advantage of simultaneously fitting the light curve and the RVs, which measures \rhostar\ (thus refining \mstar\ and \rstar).  As demonstrated below, the values for planet masses in this model agree with the planet masses determined in the circular orbital solution to within $1\sigma$, and so we adopt the parameters from the eccentric solution for the rest of this work.  

The \eke photometry and Keck radial velocities were simultaneously fit with an orbital model.  The model has the following free parameters: mean stellar density (\rhostar), scaled planetary radius ($r_n/R*$), impact parameter ($b_n$), orbital period ($P_n$), center of transit time ($T_{c,n}$), radial velocity amplitude ($K_n$), eccentricity ($e_n$) and argument of pericenter ($w_n$) via esin$w_n$ and ecos$w_n$.  A photometric and radial velocity zero point were also included.  The number (n=1,2,3,4) corresponds to the parameters for planet b,c,d and e respectively.   The transit model uses the quadratic formulae of \citet{Mandel2002}.  Limb-darkening coefficients were fixed in the models to 0.3236 and 0.3052 as determined from the grid of \citet{Claret2011}.  The orbits are modeled with non-interacting Keplerians.

A best-fit model was initially computed by minimizing $\chi^2$ with a Levenberg-Marquardt style algorithm. This model was used to measure TTVs and to seed an MCMC analysis of the model parameter space.  TTVs were determined by fitting for each individual transit, fixing all parameters except $T_c$ to their best-fit values.  An updated best-fit model was then computed using the TTVs to produce a better phased transit for each planet.  The time-series were corrected by computing time-corrections based on a linear-interpolation of the TTVs.

Posterior distributions for each model parameter were determined with an MCMC-style algorithm.  This model has been described in \citet{Gauthier2012} and \citet{Borucki2012}, the only difference is that the TTV measurements are included and fixed to their best-fit values.  Four Markov-chains were calculated each with a length of $10^6$. The first 10\% of each chain was discarded as burn-in.  The median and $\pm 1\sigma$ percentiles were calculated for each model parameter and reported in Table \ref{tab:SystemParams}, which is in the appendix due to its length.

\begin{figure}[htbp] 
   \centering
   \includegraphics{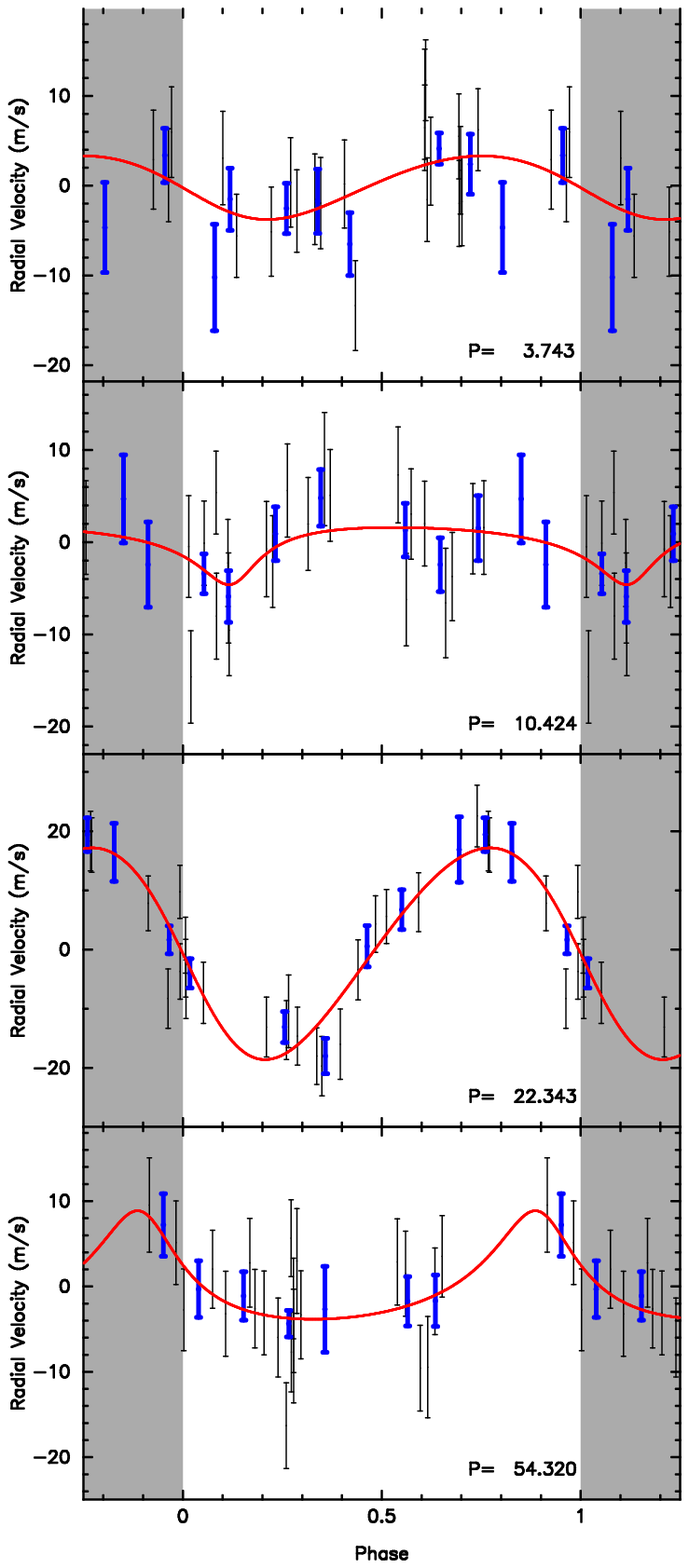}
   \caption{\footnotesize Radial velocity components from the four-planet eccentric fit in which the photometry and radial velocities were fit simultaneously.  Each panel shows the radial velocity signature from one planet (top to bottom: b, c, d, e).  The red line is the model fit; the black points are the RV data minus the model RVs from the other three planets.  The blue points are the binned RVs; their error bars the uncertainty in the mean.  The shaded regions show phase-shifted repetitions of the data and fit.}  
   \label{fig:rv_phase_ecc}
\end{figure}

\subsubsection{Stellar Properties}
We used the template spectrum (without iodine) to determine the effective temperature (\teff = \teffSME), surface gravity (\logg = \loggSME), and metallicity (\feh = \fehSME) of \starname\ through the Spectra Made Easy (SME) analysis technique described in \citet{Valenti1996}.  Applying the evolutionary constraints of the  \citet{Yi2001} model isochrones and the simultaneous solution of the lightcurve and RVs, we determine the mass ($\mstar = \mstarRowe$) and radius ($\rstar = \rstarRowe$) of \starname.  The stellar properties are presented in Table \ref{tab:SystemParams}.

\subsubsection{Properties of \planetd\ from the Eccentric Orbit Solution}
We detect \planetd\ with 9$\sigma$ confidence with the RVs.  The eccentric orbit analysis gives a mass of \mplanetd \mearth.  This mass is marginally consistent with the mass reported by \citet{Hirano2012}, who found $\mpl = 73 \pm 25$ \mearth.

\subsection{Planet Masses from Transit Timing Variations.}
\begin{figure}[htbp] 
   \centering
   \includegraphics[width=\imsize]{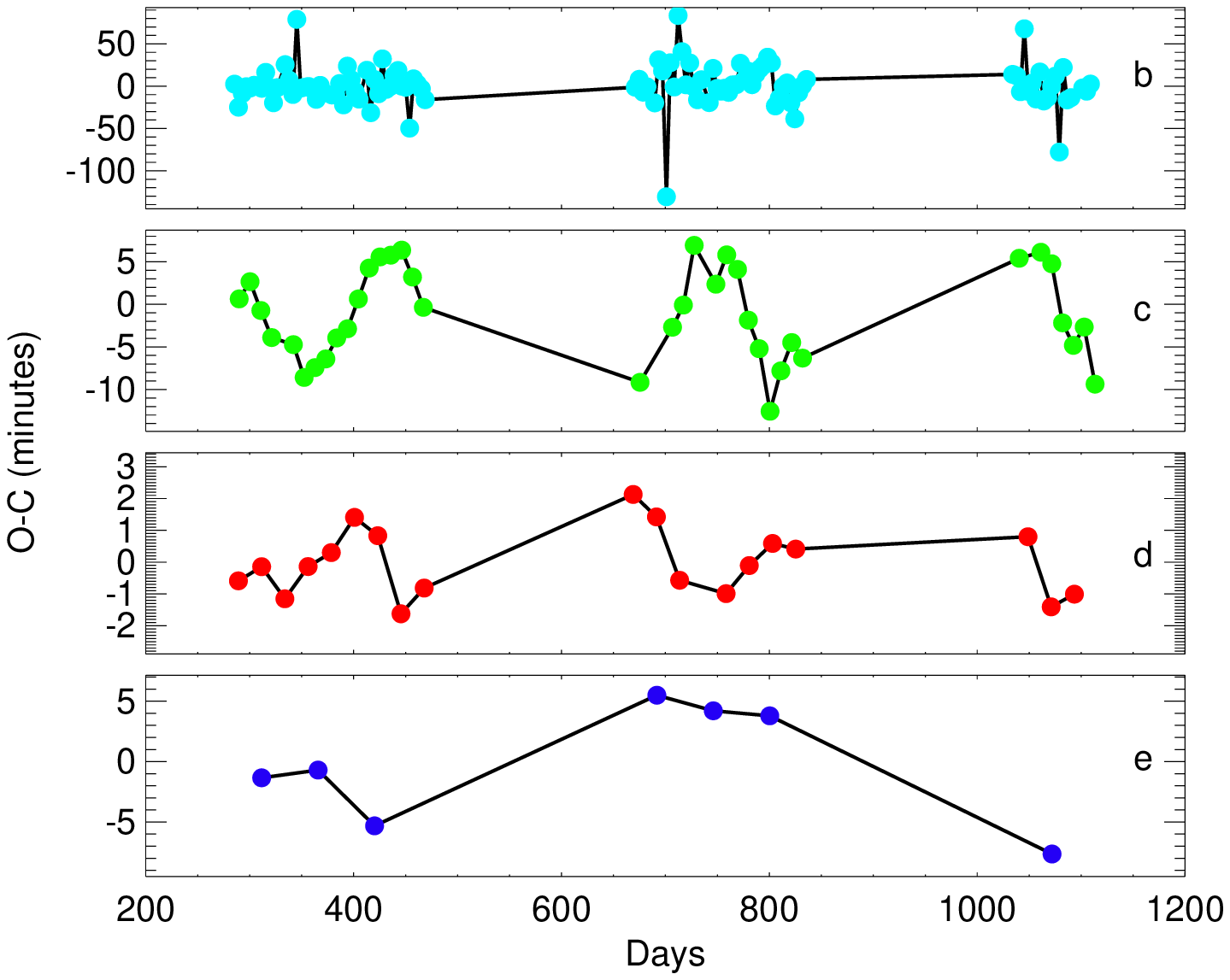} 
   \caption{\footnotesize Observed Transit Timing Variations in Q1-Q12 for (top to bottom) \planetb, \planetc, \planetd, and \planete, with the same color scheme as Figures \ref{fig:jasonrowe_lc} and \ref{fig:jasonrowe_ph}.  The y-axis of each plot shows the difference between the observed transit time (O) and the transit time expected for a periodic orbit (C).  Days are in JD$-$2454900.  The errors are smaller than the point size.  We excluded one $5.5\sigma$ outlier in the O-C measurements for planet c at 686 days.  There is a section between days 650 and 800 in which the TTVs of planets c and d are anti-correlated, indicating a possible gravitational interaction.}
      \label{fig:ttvs}
\end{figure}

We observe coherent TTVs for \planetc\ and \planetd\, which are presented in Figure \ref{fig:ttvs}.  TTVs usually indicate gravitational interactions between adjacent pairs of planets; such interactions allow us to refine the mass estimates of these planets.

We use the prescription of Equations 8 and 9 from \citet{Lithwick2012} to predict the anti-correlated TTV signals produced during the interaction of planets c and d between 650 and 800 days.  Assuming zero eccentricity for both planets and ignoring factors of order unity, we calculate

\begin{equation}
V = P \frac{\mu'}{\pi j^{(2/3)}(j-1)^{(1/3)}\Delta}
\end{equation}
\begin{equation}
V' = P' \frac{\mu}{\pi j(j-1)^{(1/3)}\Delta}
\end{equation}
where $\mu$ is the ratio of the inner planet mass to the stellar mass, $\mu'$ is the ratio of the outer planet mass to the stellar mass, $V$ is the predicted semi-amplitude of the complex TTV signal for the inner planet, and $V'$ is the predicted semi-amplitude of the complex TTV signal for the outer planet.  $\Delta$, which is given by

\begin{equation}
\Delta = \frac{P'}{P}\frac{j-1}{j}-1
\end{equation}

is the fractional departure from a $j:j-1$ mean motion resonance.

Using the values for the orbital periods of planets c and d, for which the closest first-order mean motion resonance is 2:1, we calculate $\Delta = 0.074.$  Using the SME-determined mass of the star and the RV-determined planet masses, we calculate $V = 7$ minutes and $V' = 1.7$ minutes.  These values agree with the observed TTV interaction in Figure \ref{fig:ttvs}, for which the TTV amplitudes of planets c and d appear to be about 7 and 1 minutes, respectively.

\subsection{Dynamical Stability}
We investigate the long-term stability of the KOI-94 system. We integrate the orbits of all four planets using the built-in hybrid symplectic/Bulirsch-Stoer integrator, part of the Mercury software package \citep{Chambers1999}. We use the orbital and physical parameters of the four planets detailed in Table \ref{tab:SystemParams}. The mean anomalies are derived from the best-fit joint radial-velocity solution. We define in the following a non-stable system when a close encounter between a given body and one of the four others occurs, within their common Hill radius. In a first set of 1-Myr integrations, fifteen eccentricity values for each planet are randomly drawn from the normal distribution $N(e,\sigma_e^2)$, where $e$ is the best-fit orbital eccentricity and $\sigma_e$ is the $1\sigma$ eccentricity uncertainty. Eighty percent of these integrations yield close encounters between the two innermost planets b and c. Closer inspection of the results reveal that the proximity of b and c put stringent constraints on their orbital eccentricity, only allowing  values less than or equal to the best-fit values for the system to be stable. 

In a second set of ten integrations, all planets are restricted to circular orbits with masses drawn from their normal distributions, in a similar manner as for orbital eccentricity in the previous step. All integrations with circular orbits resulted in stable systems, showing that varying planetary masses within their $1\sigma$ uncertainty has a negligible influence on the dynamical stability of the system, if orbits are kept circular. 

More detailed dynamical analyses of the KOI-94 system would determine an upper limit on orbital eccentricity for the b and c components. This would however imply arbitrary assumptions on planetary masses, which would not completely exclude configurations with large orbital eccentricities. A possible resolution would be to precisely determine the planetary masses of the close-in components, either through additional RV observations or an N-body analysis of the TTV signals from these planets.

\section{Support for the Existence of the Planets}
In this work, we sought to measure the mass of the warm Jupiter \planetd\ to determine how the planet compared to the population of hot Jupiters.  We simultaneously fit the \eke light curve and \nRV\ RVs from Keck/HIRES to measure the mass of \planetd\ with a statistical significance of $9\sigma$.  We also sought to measure the masses of the other three transiting planet candidates in \starname\ in order to better understand the architecture of this planetary system.  Our RV measurements of \starname\ measure the masses of the other planets with signficances of less than 3$\sigma$ (see Tables \ref{tab:SystemParamsCirc} and \ref{tab:SystemParams}).  However, other observations we have made support the interpretation of \starname\ as a system with four transiting planets.  We describe support for the planetary interpretation of the candidates below.  We defer our discussion of the comparison of \planetd\ to other Jupiter-size planets to the next section.  

We report the properties of planets \planetb\ (\mpl = \mplanetb \mearth, \rpl = \rplanetb \rearth), \planetc\ (\mpl = \mplanetc \mearth, \rpl = \rplanetc \rearth), \planetd\ (\mpl = \mplanetd \mearth, \rpl = \rplanetd \rearth), and \planete\ (\mpl = \mplanete \mearth, \rpl = \rplanete \rearth).  Although the fidelity of the \eke candidates is very high \citep[$90-95\%$ according to][see Fressin et al. (in prep) for an estimate of fidelity as a function of planetary radius]{Morton2011}, false positives do exist among the \eke planet candidates in the form of background eclipsing binaries, hierarchical triples, and other configurations of stars that result in the dilution of the eclipse signal, allowing it to masquerade as a planetary transit.  The large size of \eke pixels renders \eke particularly vulnerable to bound companion stars with a large planet, such as a Neptune- or Jupiter-size planet, mimicking the transit signal of an Earth-size planet.  To show that a planetary interpretation is superior to the interpretation of these various false positive signals, we outline the various measurements and statistical properties of the \starname\ system that support the hypothesis of four transiting planets.

\subsection{Radial Velocities from \planetd}
The eccentric orbit solution of the RVs and light curve yield a semi-amplitude of \semiAmpd\ \ms\ with a \perioddshort\ day period for \planetd.  Similarly, the four-planet circular orbit solution yields a semi-amplitude of \semiAmpdcirc\ \ms\ to the RVs.  The agreement between the circular and eccentric values for the semi-amplitude of \planetd\ underscores the robustness of this measurement.  

At a 22-day period, a semi-amplitude of \semiAmpd\ \ms\ is consistent with the orbit of a planet around a star.  A binary star system in 22-day period would have a velocity semi-amplitude of hundreds of kilometers per second.  The width of the spectral lines is consistent with a stellar rotation speed of 8 $\mathrm{km s^{-1}}$, which is at least an order of magnitude smaller than the orbital speed of a such a binary system.  Thus, we rule out the possibility of an eclipsing binary false positive in a 22-day orbital period for \planetd.  This supports the interpretation of \planetd\ as a planet.  As discussed below, the planetary status of \planetd\ strengthens the case that the other transiting candidates are also planets.

\subsection{Observed TTV signature between \planetc\ and \planetd}
The apparent anti-correlation in the TTVs between planets c and d from 700-800 days (see Figure \ref{fig:ttvs} suggests that these bodies are dynamically interacting, i.e. that \planetc\ is also a planet.  Our order-of-magnitude treatment of the TTV signatures indicates that the interacting bodies are of planetary masses, although the mass estimates are too low by a factor of 5.

\subsection{No Evidence of Another Star}
The adaptive optics images show no evidence of companions as close as 0.5\arcsec\ from KOI-94 within 2 magnitudes in the J band or 3.4 magnitudes in the Ks band. Similarly, the speckle imaging shows no evidence of companions as close to KOI-94 as 0.6\arcsec\ within 4 magnitudes in the R band.  Also, the $3\sigma$ upper limit on an RV trend of $\dot{\gamma} = 6.9$ \ms\ per year between fall 2009 and fall 2012 rules out a Jupiter-mass or more massive companion within 5 AU.

The detection of a second stellar spectrum in the spectrum of \starname\ would indicate background or companion star.  We searched for the spectrum of a second star in the iodine-free HIRES spectrum of \starname.  To fit the spectrum of the primary star (\starname), we used a library of over 700 observed spectra from Keck HIRES that span \teff: $3266\,\mathrm{K}-7258\,\mathrm{K}$, $\logg: 1.46-5.0$, and \feh:$ -1.47- +0.56$.  We found the spectrum from this library with the least squares difference from the spectrum of \starname\ (with similar results to the SME analysis) and subtracted this spectrum.  We then compared the residual spectrum to each spectrum in the stellar library.  The deepest minimum in $\chi^2$ between the residual spectrum and another star from the library was $2\sigma$ at $-59.2 \kms$; however, there were many other 2$\sigma$ solutions for that spectrum.  We did not detect a second stellar spectrum with $>1\%$ of the observed brightness of \starname\ and a relative radial velocity of at least $\sim8\,\kms$ with $>2\sigma$ significance.  This technique is sensitive only to neighboring stars within 0.4\arcsec, the half-width of the slit of the HIRES spectrometer.

\subsection{Low False Positive Rate in Multi-Planet Systems}
\citet{Lissauer2012} uses statistical arguments to calculate the false positive fraction in multi-planet systems.  Given the observed number of \eke targets $n_t$, the number of \eke planet candidates $n_c$, the number of \eke multi-planet systems with $i$ planet candidates $n_{m,i}$, and the planet fidelity $P$ (or single candidate false positive rate $1-P$), the fraction of systems with four planet candidates that we expect to consist of one false positive and three true planets ($\mathrm{P_{1FP}}$) is equivalent to the probability that a false positive is lined up behind a system that already has three true planets.  The number of false positives among the candidate transiting planets is $(1-P)n_c$, and the fraction of those aligned with apparent $i$-planet systems is $n_{m,i}/n_t$.  Thus,
\begin{equation}
\mathrm{P_{1FP}} = \frac{(1-P)n_c \frac{n_{m,4}}{n_t}}{n_{m,4}} =  \frac{(1-P)n_c}{n_t}.
\end{equation}
Adopting the values $n_t = 160171$ from \citet{Lissauer2012} and $n_c=2300$ from \citet{Batalha2012}, and assuming $P = 0.9$ \citep[in accordance with][]{Morton2011}, we calculate $$\mathrm{P_{1FP}} = 0.0014.$$  The probability that 2, 3, etc. planet candidates are all false positives (un-associated eclipsing binaries that all happen to align behind \starname\ within 0.5\arcsec) is orders of magnitude smaller and can be ignored.  The low false positive probability is definitive of planethood even without the other arguments presented in this section.

\subsection{Rossiter-McLaughlin Measurement During the Transit of \planetd}
\citet{Hirano2012} measured the Rossiter-McLaughlin (RM) effect during transit of \planetd.  They observed a clear RM signal that, when measured considering a transit depth of $\rpl/\rstar = 0.06856 \pm 0.00012$, implied a projected stellar rotation of $V\mathrm{sin}I_s = 8.01 \pm 0.73 \kms$, which is in good agreement with their spectroscopically determined value of $7.33 \pm 0.32 \kms$ and our SME analysis.  This constitutes evidence for the planetary status of \planetd.  \citet{Albrecht2013} also measured and modeled the RM effect during the same transit of \planetd and obtained results that agreed with \citet{Hirano2012}.

Modeling of the transit durations, including ingress and egress, of the \starname\ transit candidates indicate small inclinations with respect to the line of sight (see the inclinations in Table \ref{tab:SystemParams}.  \citet{Hirano2012} measure a projected mutual inclination between planets d and e of $\delta = -1.5^\circ$ during a mutual transit event, indicating that these bodies are coplanar.  That these planets are aligned with the stellar spin axis strengthens the argument for their planetary status.  Recent work by \citet{Fabrycky2012} shows that many \eke multi-planetary systems are coplanar, so it is likely that planets b and c are coplanar with planets d and e.

\section{Constraints on the Composition of \planetd}
The goal of this work was to measure the mass of the giant planet, KOI-94d, and determine whether its bulk density was consistent with that of an inflated or a cold Jupiter.


We have modeled the thermal evolution and contraction of planet d using the methods described in \citet{Fortney07} and \citet{Miller11}.  This model assumes no extra heat source from the star.  Including uncertainties in the system age, planet mass, planet radius, orbital semi-major axis, and heavy element distribution within the planet \citep[see][]{Miller11} we estimate that $18 \pm 6$ \mearth\ of heavy elements are contained within the planet.  This is very similar to estimates for Saturn \citep{Saumon04}.  Based on the \feh\ of the parent star determined from our SME analysis of the spectrum, we estimate that the metals mass-fraction of planet d ($Z_{\rm planet}$) is $11 \pm 4$ times that of the parent star.  This metallicity enhancement, at this planet mass, agrees well with other ``warm Jupiter" planets studied by \citet{Miller11}.

We use state of the art thermal evolution models for giant planets to establish that the bulk density of the planet is fully consistent with a non-inflated planet, and it is indeed enhanced in heavy elements compared to its parent star, in a manner similar to Saturn.  We can furthermore compare KOI-94d to other planets by creating mass-radius and mass-density plots for all planets with measured radius and mass (see Figure \ref{fig:mass-density}). These plots demonstrate the dependence of planetary radius (and density) on mass and incident flux.

\section{The Radius-Mass-Incident Flux Relation}
In this section, we examine empirical relations between the radius, mass and incident flux of exoplanets, including the \starname\ system.  We discuss possible physical interpretations of these relations and suggest avenues of future theoretical investigation.

\begin{figure*}[htbp] 
   \centering
   \includegraphics[width=2.8in]{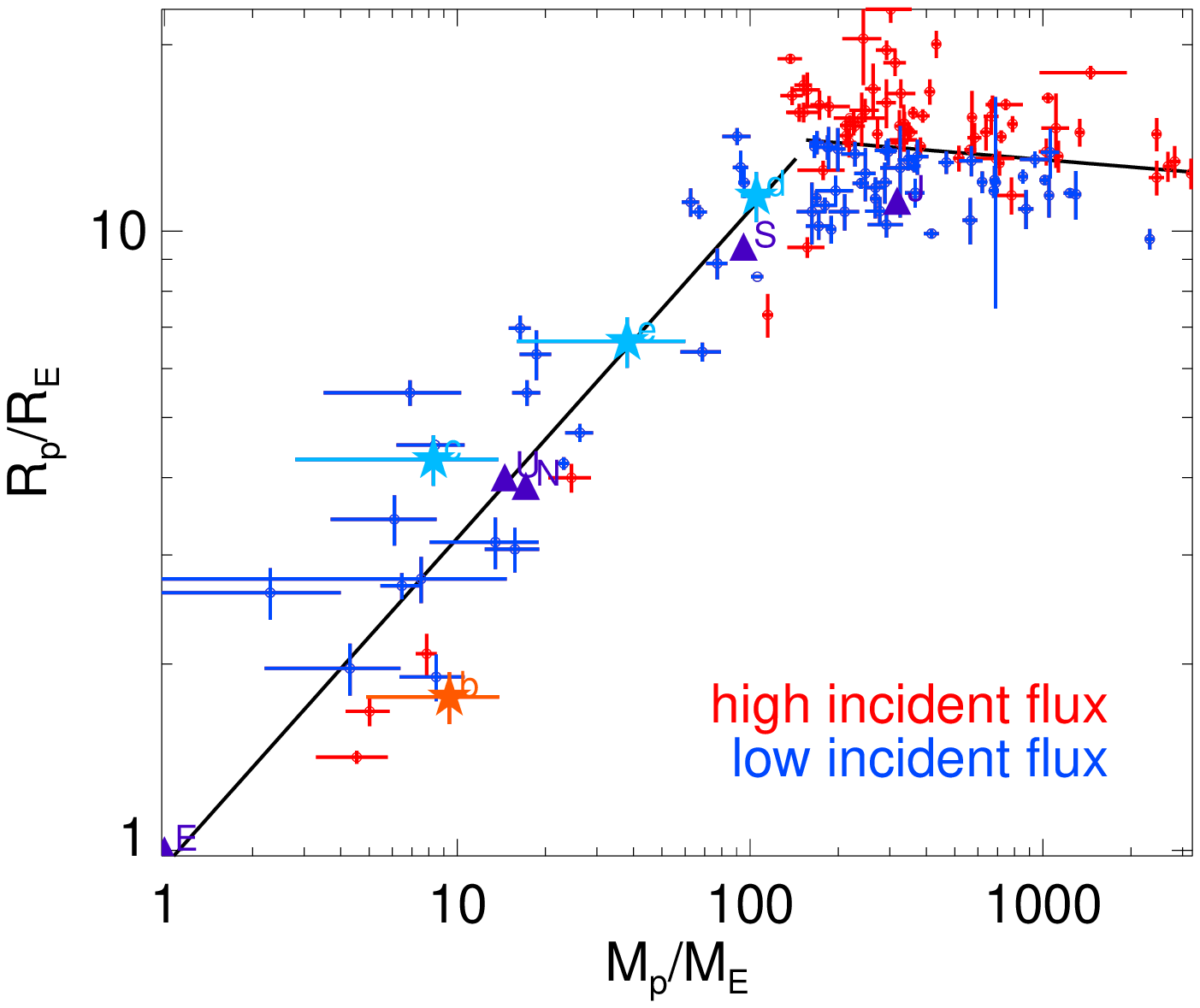} 
   \includegraphics[width=2.8in]{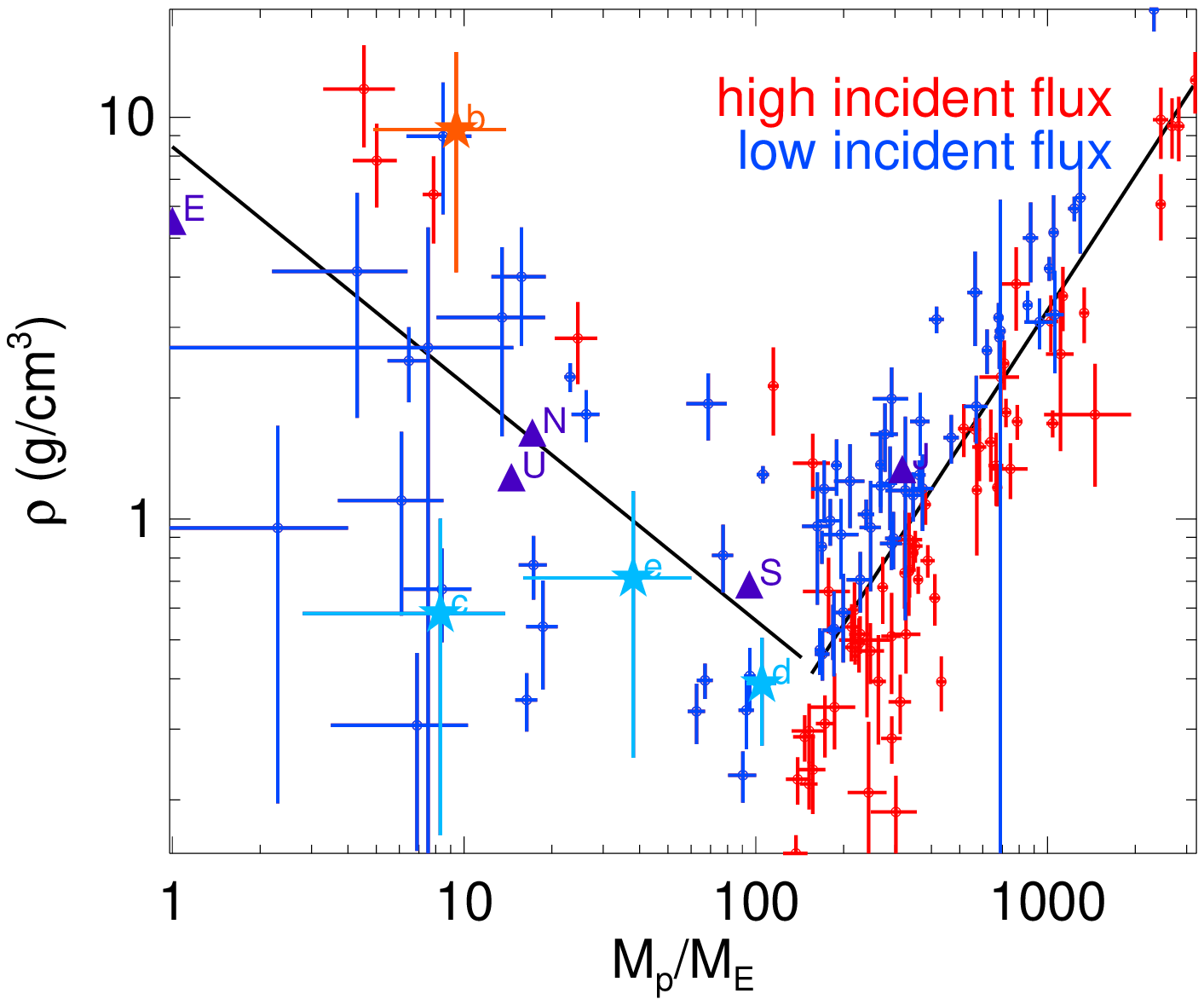} 
   \caption{\footnotesize All planets with measured mass, radius, incident flux, and uncertainties therein, as listed in exoplanets.org \citep[see Table \ref{tab:all_planets}]{Wright2011}.  KOI-94 planets are plotted as five-pointed stars; Solar system planets are plotted as triangles.  \textbf{Left:}  Planet radius vs. planet mass. We divide the planets into two populations: those with higher-than-median incident flux (red), and those with lower-than-median incident flux (blue)   The Solar system planets (purple) all receive less than the median incident flux; KOI-94 planets c, d, and e (cyan) receive less than the median incident flux, while \planetb\ (orange) receives more than the median incident flux.  For $\mpl > \mspecial$, higher incident flux correlates with larger planetary radius.  For $\mpl < \mspecial$, higher incident flux ($F$) correlates with smaller planetary radius.  The best-fit planes \forsolid\ and \forgiant\ are shown at the median flux $F = \medianflux\ \fluxunit$.  \textbf{Right:}  Planet density vs. planet mass.  The coloring is the same as in the left panel, and the density fits \forsolid and \forgiant are shown at the median flux.  For $\mpl > \mspecial$, higher incident flux correlates with lower bulk density.  For $\mpl < \mspecial$, higher incident flux correlates with higher bulk density.  We determine empirical relations (see text) between log(\mpl), log($F$), and each log(\rpl) and log(\rhopl) for $\mpl > \mspecial$ and $\mpl < \mspecial$.}
   \label{fig:mass-density}
\end{figure*}

This population of planets was compiled from exoplanets.org \citep{Wright2011}, which was queried on September 27, 2012.  Our selection criteria were that the mass and radius of the planet were measured, and that errors in these measurements were reported, that the effective temperature and radius of the star were measured with errors reported, and that the semi-major axis was measured with errors reported.  This resulted in the sample of 138 exoplanets listed in Table \ref{tab:all_planets}.

The methods for determining the radius, mass, and incident flux of the planets in Table \ref{tab:all_planets} were as follows:  All the planets transit, and so their radius measurement was based on transit depth as determined within a self-consistent model of the observed light curve.  For many of these planets, the uncertainty in planet radius is dominated by the uncertainty in stellar radius.  The masses of the planets were measured by one of two methods: the majority were measured as \mpl sin$i$ based on the radial velocity of the star; however, in several \eke systems of multiple planets, TTVs aided \citep[Kepler-18]{Cochran2011} or provided the sole means \citep[Kepler-11]{Lissauer2011} of planetary mass calculation.  Uncertainties in the planet mass stem from uncertainties in the stellar mass, uncertainties in the RV semi-amplitude $K$ for low-mass planets, and uncertainties in the TTV analysis.  The incident flux for each planet was calculated using Equation \ref{eqn:incident_flux}, so uncertainties in the incident flux relate to uncertainties in stellar effective temperature, stellar radius, and semi-major axis.

There is a break in the mass-radius relation at $\sim$\mspecial\ (see Figure \ref{fig:mass-density}).  We chose \mspecial\ as the break based on a visual inspection of the mass-radius and mass-density diagrams. In determining the relation between planet radius, planet mass, and incident flux, we consider planets more or less massive than \mspecial\ separately.  Our sample included 35 planets with \mpl $<$ \mspecial\ and 103 planets with \mpl $>$ \mspecial.  All four planets in the \starname\ system are included in the low-mass population.

To determine how incident flux affects radius, we calculated the time-averaged incident flux on each exoplanet from Equation \ref{eqn:incident_flux}.  We divided the population into the ``high incident flux" half (those with incident fluxes larger than the median incident flux, $F_m = 8.6 \times 10^8 \fluxunit$), and the ``low incident flux" half.  These are shown in Figure \ref{fig:mass-density} as the red (high flux) and blue (low flux) sets of points.  For planets with $\mpl > \mspecial$, the planets that receive high incident flux are systematically larger than planets that receive low incident flux.

\begin{figure}[htbp] 
   \centering
   \includegraphics[width=\imsize]{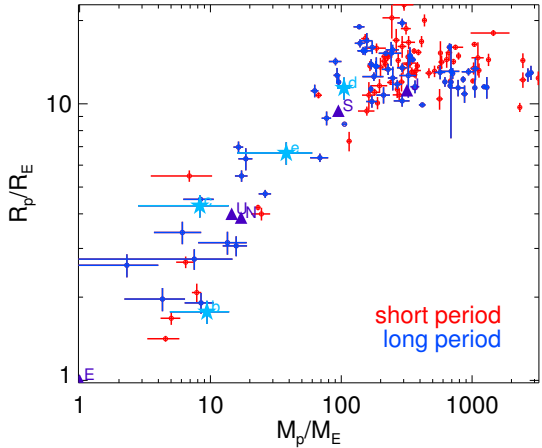} 
   \caption{\footnotesize  Planet radius vs. planet mass (same Figure \ref{fig:mass-density}, but with a different coloring scheme).  We divide the planets into those with lower-than-median orbital periods (red) and those with higher-than-median orbital periods (blue).  The Solar System planets (purple) and KOI-94 planets b, c, d, and e (cyan) all have longer than the median period.  Orbital period does not correlate with planet radius.}
   \label{fig:mass-radius-period}
\end{figure}

We performed a similar test to determine how the orbital period affects radius (see Figure \ref{fig:mass-radius-period}).  We divided the exoplanet population into those with lower than the median orbital period of 3.52 days (red, ``short period") and higher than the median orbital period (blue, ``long period").  Planet radius does not correlate with orbital period.

Using the KOI-94 system and all other exoplanets with published values for both mass and radius, we establish two fundamental planes for exoplanets that relate their mass, incident flux, and radius from a few Earth masses up to ten Jupiter masses.  We fit two planes between log(\rpl), log(\mpl), and log($F$), one in each mass regime.  The resulting relations are:
\begin{equation}\label{eqn:plane1}
\begin{split}
\frac{\rpl}{\rearth} = \rcoeffs \left(\frac{\mpl}{\mearth}\right)^{\mcoeffs} \left(\frac{F}{\fluxunit}\right)^{\fcoeffs}\\
\forsolid
\end{split}
\end{equation}
and
\begin{equation}\label{eqn:plane2}
\begin{split}
\frac{\rpl}{\rearth} = \rcoeffg \left(\frac{\mpl}{\mearth}\right)^{\mcoeffg} \left(\frac{F}{\fluxunit}\right)^{\fcoeffg}\\
\forgiant
\end{split}
\end{equation}

For completeness, we also fit two planes between log(\rhopl), log(\mpl), and log($F$):
\begin{equation}\label{eqn:plane3}
\begin{split}
\frac{\rhopl}{\gcc} = \dcoeffs \left(\frac{\mpl}{\mearth}\right)^{\mdcoeffs} \left(\frac{F}{\fluxunit}\right)^{\fdcoeffs}\\
\forsolid
\end{split}
\end{equation}
and
\begin{equation}\label{eqn:plane4}
\begin{split}
\frac{\rhopl}{\gcc} = \dcoeffg \left(\frac{\mpl}{\mearth}\right)^{\mdcoeffg}\left(\frac{F}{\fluxunit}\right)^{\fdcoeffg}\\
\forgiant
\end{split}
\end{equation}

Table \ref{tab:all_planets} lists the mass, radius, incident flux, orbital period, and reference for the planets used to calculate these fits.  We include the Solar system planets in Figure \ref{fig:mass-density} for reference, although these planets were not used to generate the fits in Equations \ref{eqn:plane1}$-$\ref{eqn:plane4}.  Slice through the planes at the median incident flux is shown as black lines in Figure \ref{fig:mass-density}.

These planes were calculated by fitting a plane to the measurements of log\mpl, log\rpl, and log$F$ in each mass regime, with equal weight for each point.  To see how the uncertainties in \mpl, \rpl, and $F$ influenced the fit, we did 1000 trials in which we varied each measurement of \mpl, \rpl, and $F$ based on a Gaussian distribution with the $1\sigma$ uncertainties reported for that planet to create a posterior distribution of coefficients.  The median and average values for the posterior distribution of the coefficients were consistent with the coefficients of the original fit.  If we write Equations \ref{eqn:plane1} and \ref{eqn:plane2} more generally as $\mathrm{log}(\rpl/\rearth) = A + B \mathrm{log}(\mpl/\mearth) + C \mathrm{log}(F/\fluxunit)$, the $1\sigma$ uncertainties in the coefficients for Equation \ref{eqn:plane1} (i.e. \forsolid) are $A = 0.25 \pm 0.185$, $B = \mcoeffs \pm 0.052$, and $C = \fcoeffs \pm 0.017$.  The $1\sigma$ uncertainties in the coefficients for Equation \ref{eqn:plane2} (i.e. \forgiant) were $A = 0.39 \pm 0.053$, $B = \mcoeffg \pm 0.0096$, and $C = \fcoeffg \pm 0.0055$.  Thus, the dependence of radius on mass for low-mass planets is significant at $10\sigma$, and the dependence of radius of flux for high-mass planets is significant at $17\sigma$.  The downward slope of radius versus mass for giant planets is detected at a significance of $4\sigma$, and the downward slope of radius versus flux for small planets is uncertain.

For $\mpl < \mspecial$, the RMS scatter of the radius is \rmsrs \rearth\ and the RMS scatter of the density is \rmsrhos \gcc.  Considering that the average radius of a planet in this mass regime is 6.72\rearth\ and the average uncertainty in planet radius in this mass regime is 0.34\rearth (i.e. 5\% of the typical planet radius), the RMS scatter of radii for $\mpl < \mspecial$ is large compared to the uncertainties in measurements of planet radii.  For $\mpl > \mspecial$, the RMS scatter of the radius is \rmsrg \rearth\ and the RMS scatter of the density is \rmsrhog \gcc.  Considering that the average uncertainty in planet radius \forgiant\ is 0.76\rearth, the RMS scatter is comparable to the uncertainty in planet radius in this mass regime.  Interpretations of the RMS scatter in each mass regime is discussed in \S7.1.

Note than \forsolid, radius depends strongly on mass ($\rpl \propto \mpl^{0.52}$) and very weakly on incident flux ($\rpl \propto F^{-0.03}$).  For $\mpl > \mspecial$, the dependence is reversed: $\rpl \propto \mpl^{-0.04}$, and $\rpl \propto F^{0.09}$.  Since mass has little effect on radius for giant planets, the incident flux is the most important factor in predicting planet radius.  

\begin{figure}[htbp] 
   \centering
   \includegraphics[width=\imsize]{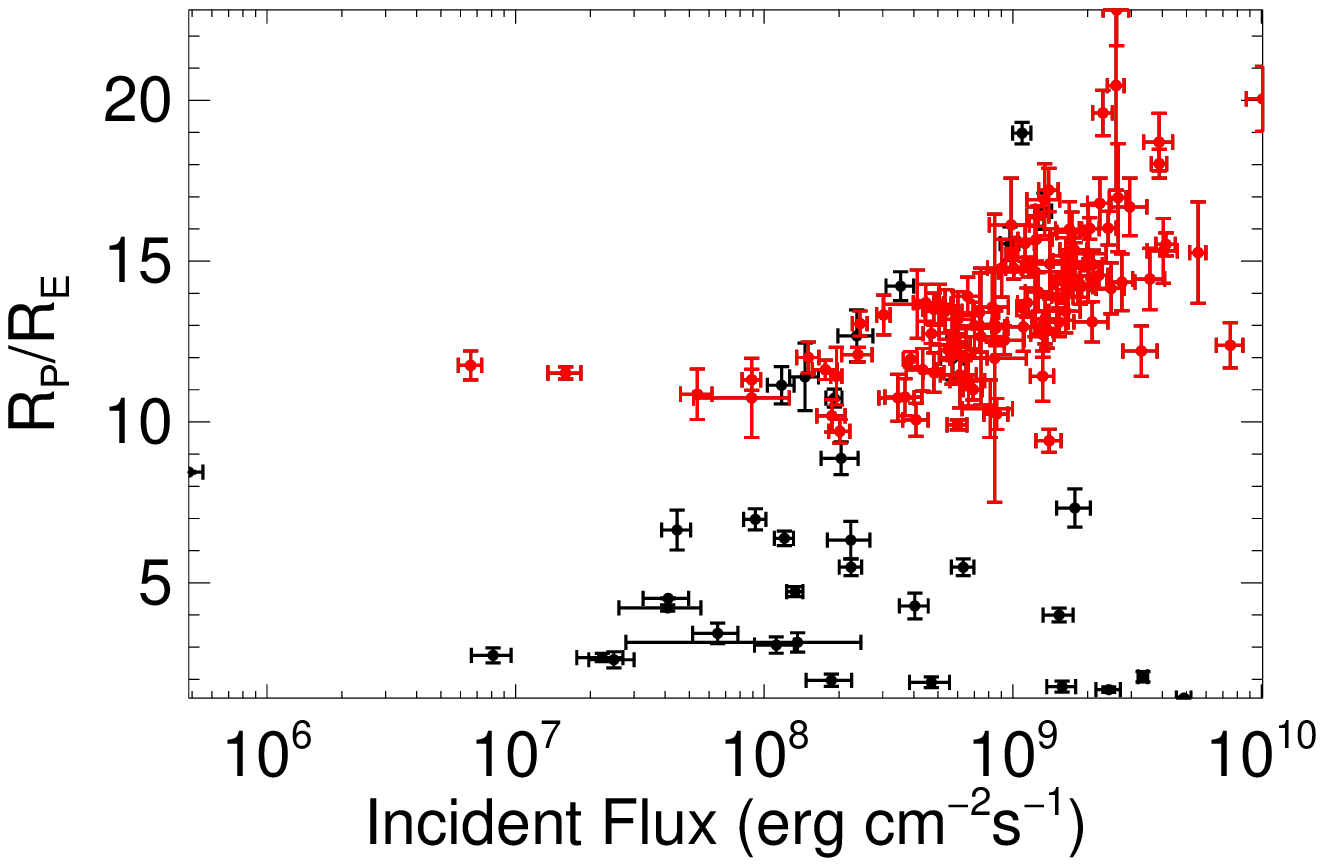} 
   \includegraphics[width=\imsize]{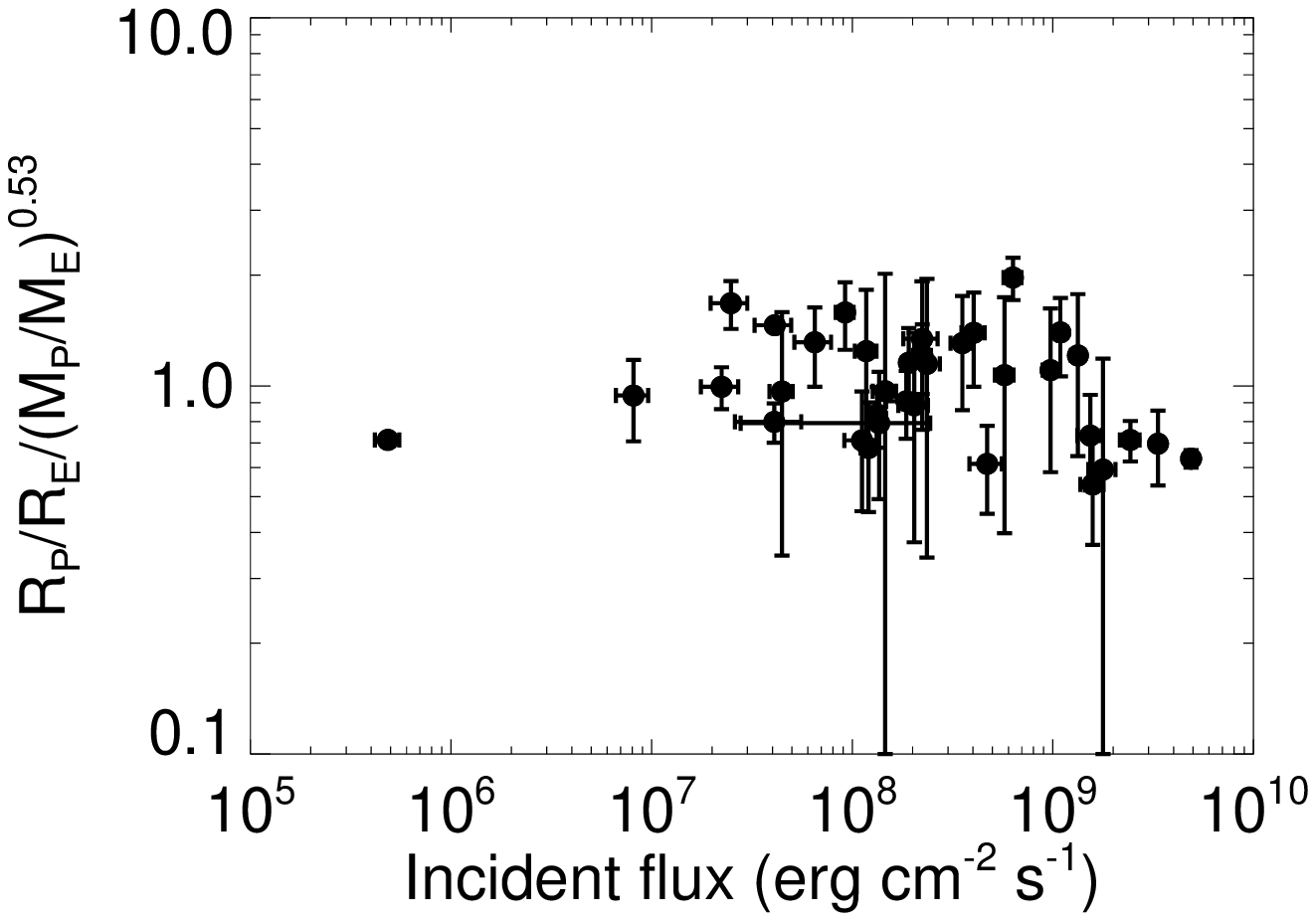}
   \caption{\footnotesize \textbf{Top:}  All planets with measured mass, radius, and incident flux, as listed in exoplanets.org (see Table \ref{tab:all_planets}).  The black points are planets with $\mpl < \mspecial$, and the red points are planets with \mpl $>$ \mspecial.  For the giant planets, planet radius increases with incident flux.  For low-mass planets, incident flux does not correlate with planet radius.  This is because planet radius scales with planet mass more strongly than with incident flux for low-mass planets, whereas incident flux is the primary factor in determining the radii of high-mass planets (see Equation \ref{eqn:plane2}).  \textbf{Bottom:} Planet radius, divided by dependence on planet mass according to Equation \ref{eqn:plane1}, versus incident flux \forsolid.  The best fit to the data suggests a slight trend toward lower radius at higher incident flux, although the data are also consistent with no correlation between incident flux and planet radius.}
   \label{fig:radius-flux}
\end{figure}

In light of the very clear dependence of giant planet radius on incident flux, and the possibility of a dependence of low-mass planet radius on incident flux, we wanted to examine the relations between incident flux and planet radius in greater detail.  The top panel of Figure \ref{fig:radius-flux} shows planet radius as a function of incident flux for the low-mass and high-mass planets.  The scatter in radius of the low-mass planets can be attributed to the strong dependence of planetary radius on mass \forsolid.  However, the relation between the radii of giant planets and the incident flux is clear in this plot because the dependence of planet radius on mass is very small \forgiant.

It appears that for planets with $\mpl < \mspecial$, the planets receiving high incident flux are systematically smaller than planets receiving low incident flux.  To examine the validity of this correlation, we plotted the residuals to the relation $\rpl \propto \mpl^{0.52}$ as a function of incident flux (see the bottom panel of Figure \ref{fig:radius-flux}).  We found that the residuals only weakly depend on incident flux, but that there is a visible downward trend.  Thus the suggestion that low-mass planets with high incident  flux are smaller remains unclear.  Future characterization of low-mass planets receiving varying amounts of incident flux will help elucidate this relation, if it exists.

\subsection{Interpretation of the Radius-Mass-Incident Flux Relations}

We consider the KOI-94 system in the context of other planets with published values for both mass and radius.  We quantify these trends in two fundamental planes (see Equations \ref{eqn:plane1}$-$\ref{eqn:plane2}) for exoplanets that relate their mass, incident flux, and radius from a few Earth masses up to ten Jupiter masses.  These equations demonstrate that for low-mass planets, mass is much more important than flux in predicting a planet's radius, whereas for high-mass planets, incident flux is more important for predicting the radius of a giant planet than the planet mass.  

For low-mass planets, inverting Equation \ref{eqn:plane1} predicts the planet's mass given its radius and incident flux.  The small coefficient for mass \forgiant\ indicates that it is difficult to predict the mass of a giant planet given its radius and incident flux.

The small RMS scatter in radius for giant planets ($\delta \rpl = \rmsrg$ \rearth) compared to the typical uncertainty in radius ($\sigma_{\rpl} = 0.76\rearth$) suggests that our model of mass and incident flux affecting the radius of giant planets is appropriate and that other factors, such as orbital period and metallicity, play a small part, if any.  

For low-mass planets, the high RMS scatter ($\delta \rpl = \rmsrs$ \rearth) compared to uncertainties in the radius ($\sigma_{\rpl} = 0.34\rearth$) indicates that additional physics might play a role in determining radius.  In particular, the composition of low-mass planets could strongly affect planetary radius.  Planets on the low-mass branch could have compositions ranging from mostly hydrogen/helium (e.g. Saturn) to mostly water \citep[e.g. GJ 1214 b]{Charbonneau2009} to mostly rock \citep[e.g. Kepler 10b]{Batalha2011}.  The equations of state of these materials are quite different, allowing a planet of 10\mearth\ to vary by a factor of 5 in radius depending on its composition, in theory.  Despite the potentially large range in compositions at a given mass, the low-mass fundamental plane works reasonably well: extrapolating this relation to Earth, we predict $\rpl =1.15 \rearth$.  

Planetary composition might be described in part by the mass-radius relation.  For low-mass planets, the exponential dependence of \rpl\ on \mpl\ is higher than expected.  For a body of constant density (for instance, a rocky planet, if we ignore compressibility), we expect $\rpl \propto \mpl^{1/3}$.  For low-mass planets, the observed relation, $\rpl \propto \mpl^{0.52}$, is steeper.  This steep increase in radius with mass cannot be explained by the compressibility of material, since compressibility would cause less increase in radius with increasing mass.  The extra increase in radius per unit mass suggests a compositional gradient. Within the low-mass regime ($\mpl < \mspecial$), higher-mass planets might have an increased admixture of volatiles.  This is supported by observations in our own solar system; Uranus, Neptune and Saturn have a larger fraction of volatiles than Earth and Venus.  

In the giant planet population (\mpl $>$ \mspecial), the decline in radius with increasing mass corresponds to the onset of electron degeneracy as an important component of the planet's pressure along with Coulomb forces.  For a body supported by electron degeneracy pressure, we expect \rpl\ $\propto$ \mpl$^{-1/3}$.  The dependence of \rpl\ $\propto$ \mpl$^{-0.04}$, which is measured with a significance of $4\sigma$, indicates that Coulomb forces still play a significant role in supporting high-mass planets.

In planetary modeling, \mspecial\ occurs where the mass-radius relation for model planets \citep[e.g.][]{Fortney07} begins to gradually flatten out.  It is marked by waning relative importance of electrostatic forces which alone lead to $R \sim M^{1/3}$, and the gradual onset of degeneracy pressure, which for complete degeneracy leads to $R \sim M^{-1/3}$.  In an approximate way, \mspecial\ can be thought of as the start of the broad maximum in this curve that leads to radius being nearly independent of mass for giant planets and brown dwarfs \citep{Zapolsky69}.

The reversal of the correlation between $\rpl$ and $F$ at \mspecial\ is an interesting feature of the exoplanet population.  The current population of observed planets can be sculpted to some degree by evaporative mass loss \citep{Baraffe04, Baraffe06,Hubbard07,Lopez2012}. X-ray and UV (XUV) photons can drive mass-loss of hydrogen/helium planetary atmospheres, and in the energy-limited escape model \citep[e.g.,][]{Erkaev2007}, the mass loss rate depends inversely on the planet density and linearly with the incident XUV flux.  Planets with masses near $\sim$\mspecial\ are those that have low bulk densities (see Figure \ref{fig:mass-density}) are and generally more susceptible to atmospheric mass loss.  Under extreme XUV irradiation, some rare planets may migrate from just above \mspecial\ to below.

Unlike the incident flux on a planet, the orbital period of a planet does not correlate with planet radius.  This suggests that the mechanism that maintains the inflated radii of giant planets is driven by the incident flux rather than by tidal forces via eccentricity damping. However, because the eccentricities of many planets are poorly constrained, it is difficult to calculate the tidal power deposited in the planet.  The uncertainty in the heat dissipation timescale further complicates our analysis, since we cannot determine how long ago various planets might have been in sufficiently eccentric orbits for tidal forcing to inflate them.  Regardless, a planet's radius is more strongly correlated with the incident flux it receives than its orbital period.  Future studies of warm Jupiters and hot Jupiters with various orbital periods and eccentricities will help elucidate the role of potential interior heating mechanisms.  However, the goodness of fit between the radius, mass, and incident flux for giant planets suggests that the current role of tidal heating (or any other inflation mechanism that is not driven by incident flux) is quite small.

There are only 9 low-mass planets out of 35 that receive more than the median incident flux and 9 that have shorter than the median orbital period, whereas the population of giant planets is more evenly split.  \citet{Howard2012} find that the occurrence of giant planets is smaller than the occurrence of small planets at orbital periods less than 10 days, suggesting that the hot Jupiters in this work are over-represented.  The over-abundance of hot Jupiters compared to hot Neptunes in this sample could be due to historic observational bias of hot Jupiters, since the detection of Neptune-size and smaller planets from the ground was infrequent before the \eke Mission.  Measuring the masses of Neptune-size and smaller planets that receive high incident flux is necessary to probe the radius-mass-incident flux relation for low-mass planets.  

\subsection{Comparison to Previous Work}
The idea of searching for correlations between planetary radius, mass, incident flux, and other measurable planetary and stellar parameters is not novel; as mentioned in \S1, \citet{Enoch2012} and \citet{Kane2012} sought empirical relations between the properties of planets and their host stars.  Here, we incorporate an additional year's worth of planet discoveries and mass determinations, especially for Kepler planets.  There are additional differences between our work and theirs, and we highlight how this study differs from previous work.

The primary difference between \citet{Enoch2012} and the work presented here is that \citet{Enoch2012} study 16 planets within orbital periods of 10 days and with masses between 0.1 and 0.5 \mjup, whereas here we study 35 planets with masses below 150 \mearth\ (0.5 \mjup).  Whereas \citet{Enoch2012} consider three mass regimes of planets ($0.1 < \mpl < 0.5 \mjup$, $0.5 \mjup < \mpl < 2.0\mjup$, $2.0 \mjup < \mpl < 12 \mjup$), here, we only consider two regimes ($\mpl < \mspecial$, $\mspecial < \mpl < 13 \mjup$).  With the additional transiting planets included here, we do not see evidence for three mass domains.

However, we find that fits to the data incorporating only mass and incident flux predict the radius of a planet as well as the fits described in \citet{Enoch2012}.  \citet{Enoch2012} find an average absolute deviation in the predicted radius from the true radius of 0.11 \rjup\ (1.23 \rearth) across all three of their mass regimes, whereas we find a mean absolute difference of \rmsrs\ \forsolid and \rmsrg\ \forgiant, or 1.23 \rearth, over the whole sample.  Incorporating the orbital period of the planet, the stellar metallicity, and the stellar age does not significantly improve the accuracy of the predicted planetary radius.  

The slope of the mass-radius fit for $\mpl < \mspecial$ reported here, $\mcoeffs \pm 0.0052$, is only 1.5$\sigma$ different from the value of 0.45 reported in \citet{Enoch2012} for planets with $\mpl < 0.5\mjup$.  Our slope of the mass-radius fit for the high-mass planets, \mcoeffg, falls between the values for middle- and high-mass regimes in \citet{Enoch2012}, 0 and -0.09.

\citet{Kane2012} used a similar prescription to our method to obtain a mass-radius relation.  They fit a power law between the mass and radius of low-mass planets.  However, they assumed that giant planet radius was constant with planet mass (at one Jupiter radius).  Furthermore, they did not consider the effects of incident flux on planetary radius in either mass regime.

\section{Conclusions}
\begin{enumerate}
\item{\textbf{The \starname\ system}}
In this paper, we presented \nRV\ radial velocity measurements of KOI-94 obtained on Keck/HIRES.  These measurements confirm the giant planet \planetd\ and strongly support the existence of other transiting planets in the system.
\begin{enumerate}[(a)]
\item{\textbf{Properties of Planet \planetd}}
The mass is \mplanetd \mearth.  The radius is \rplanetd \rearth.  The density is \rhoplanetd \gcmc.  The planet is enriched in metals by a factor of $11 \pm 4$ with respect to the parent star.  The mass of heavy elements, or ``metals," in the planet is $18 \pm 6$ \mearth.
\item{\textbf{Properties of Planets \planetb, \planetc, \planete}}
These planets were detected at significance of less than $3\sigma$ in the radial velocity data.  The radial velocity detections of planets e and b are marginal ($> 2\sigma$), whereas the radial velocity measurements of planet c are consistent with a non-detection to $1\sigma$.  More RVs and a numerical analysis of the TTVs are needed to better characterize these planets.    However, the TTVs, multiplicity of the system, lack of evidence for another star, RM effect during the transit of \planetd\, and coplanarity of these objects strongly suggest that these candidates are planets.
\item{\textbf{Dynamical Stability of the \starname\ System}}
The system is dynamically stable on a 1 Myr timescale for a variety of configurations, including circular orbital solutions.  Although some eccentric solutions are stable, the best-fit solution considered in this work is unstable due to close encounters of planets b and c.
\item{\textbf{No Massive Outer Perturbers}}
The non-detection of a trend in radial velocity, the lack of spectral features from a second star to 1\% of the brightness of \starname within 0.4\arcsec, and the non-detection of a companion outside 0.5\arcsec (in both AO and speckle imaging) rule out large, outer companions.
\end{enumerate}
\item {\textbf{Radius-Mass-Incident Flux Relation}} Using the KOI-94 system and other exoplanets (138 exoplanets total) with published values and uncertainties for planet mass, radius, and incident flux, we establish two fundamental planes for exoplanets that relate planet radius, planet mass, and incident flux between 2 and 3000 \mearth\ in Equations \ref{eqn:plane1}$-$\ref{eqn:plane4}.  The slope of the mass-radius relation for low-mass planets suggests that as low-mass planets increase in mass, the admixture of volatiles increases.  Although the plane for \forgiant\ fits the giant planets very well, a higher RMS compared to uncertainties in radius ($~400\%$) for the low-mass planets suggests that additional physics, such as the composition of heavy elements, might contribute to the radii of these planets.
\end{enumerate}


\section*{Acknowledgments}
This project was possible thanks to NASA's \eke Mission, which provided the photometry of KOI-94 and identified it as a target worthy of follow-up.  The authors thank the \eke Team as a whole, and also a few individuals for their specific roles:  Bill Borucki and David Koch for designing the mission, and Natalie Batalha for leadership in the scientific analysis of the \eke planetary candidates.

LMW thanks Yoram Lithwick for helpful conversations and assistance in interpreting the TTV data.

LMW is financially supported by the National Science Foundation Graduate Research Fellowship Program, Grant DGE 1106400.

Spectra of \starname\ were obtained at the W.~M.~Keck Observatory.  The authors wish to extend special thanks to those of Hawai`ian ancestry on whose sacred mountain of Mauna Kea we are privileged to be guests.  Without their generous hospitality, the Keck observations presented herein would not have been possible.



\bibliography{koi94_arxiv}{}
\bibliographystyle{apj}

\appendix
\section{Long Tables}
\LongTables

\begin{deluxetable*}{lcc}
\tabletypesize{\small }
\tablewidth{0pc}
\tablecaption{Star and planet parameters for the \starname\ system.\label{tab:SystemParams}}
\tablehead{\colhead{Parameter}	& 
\colhead{Value} 		& 
\colhead{Notes}}
\startdata
\sidehead{\em Transit and orbital parameters: \planetb}
Orbital period $P$ (days)			& \periodb		& A	\\
Midtransit time $E$ (BJD)			& \epochb		& A	\\
Scaled semimajor axis $a/\rstar$		& \scaledSemiMajb	& A	\\
Scaled planet radius \rpl/\rstar		& \scaledPlanetRadiusb	& A	\\
Impact parameter $b$                		& \impactb   		& A 	\\
Orbital inclination $i$ (deg)			& \inclinationb		& A	\\
Orbital semi-amplitude $K$ (\ms)		& \semiAmpb		& B	\\
Orbital eccentricity $e$			& \eccb 		& B	\\
Center-of-mass velocity $\gamma$ (\ms)		& \gammaVelb		& B	\\

\sidehead{\em Transit and orbital parameters: \planetc}
Orbital period $P$ (days)			& \periodc		& A	\\
Midtransit time $E$ (BJD)			& \epochc	& A	\\
Scaled semimajor axis $a/\rstar$		& \scaledSemiMajc	& A	\\
Scaled planet radius \rpl/\rstar		& \scaledPlanetRadiusc	& A	\\
Impact parameter $b$                		& \impactc   		& A 	\\
Orbital inclination $i$ (deg)			& \inclinationc		& A	\\
Orbital semi-amplitude $K$ (\ms)		& \semiAmpc		& B	\\
Orbital eccentricity $e$			& \eccc 		& B	\\

\sidehead{\em Transit and orbital parameters: \planetd}
Orbital period $P$ (days)			& \periodd		& A	\\
Midtransit time $E$ (BJD)			& \epochd	& A	\\
Scaled semimajor axis $a/\rstar$		& \scaledSemiMajd	& A	\\
Scaled planet radius \rpl/\rstar		& \scaledPlanetRadiusd	& A	\\
Impact parameter $b$                		& \impactd   		& A 	\\
Orbital inclination $i$ (deg)			& \inclinationd		& A	\\
Orbital semi-amplitude $K$ (\ms)		& \semiAmpd		& B	\\
Orbital eccentricity $e$			& \eccd 		& B	\\

\sidehead{\em Transit and orbital parameters: \planete}
Orbital period $P$ (days)			& \periode		& A	\\
Midtransit time $E$ (BJD)			& \epoche		& A	\\
Scaled semimajor axis $a/\rstar$		& \scaledSemiMaje	& A	\\
Scaled planet radius \rpl/\rstar		& \scaledPlanetRadiuse	& A	\\
Impact parameter $b$                		& \impacte   		& A 	\\
Orbital inclination $i$ (deg)			& \inclinatione		& A	\\
Orbital semi-amplitude $K$ (\ms)		& \semiAmpe		& B	\\
Orbital eccentricity $e$			& \ecce 		& B	\\

\sidehead{\em Observed stellar parameters}
Effective temperature \teff (K)			& \teffSME		& C 	\\
Spectroscopic gravity \logg (cgs)		& \loggSME		& C	\\
Metallicity \feh				& \fehSME		& C	\\
Projected rotation \vsini (\kms)		& \vsiniSME		& C	\\

\sidehead{\em Fundamental Stellar Properties}
Mass \mstar (\msun)				& \mstarRowe		& D	\\
Radius \rstar (\rsun)  				& \rstarRowe		& D	\\
Surface gravity \loggstar\ (cgs)		& \loggSME		& D	\\
Luminosity \lstar\ (\lsun)			& \lumstarRowe		& D	\\
\eke Magnitude $K_p$ (mag)		& \KepMag   		& D	\\
Age (Gyr)					& \ageRowe		& D	\\

\sidehead{\em Planetary parameters: \planetb}
Mass \mpl\ (\mearth)				& \mplanetb		& B,C,D	\\
Radius \rpl\ (\rearth)				& \rplanetb		& A,B,C,D	\\
Density \rhopl\ (\gcmc)				& \rhoplanetb		& A,B,C,D	\\
Orbital semi-major axis $a$ (AU)			& \semiMajb		& E		\\
Incident Flux $F$ (\fluxunit)			& $1.58 \times 10^9$    &A,C \\
Equilibrium temperature \teq\ (K)		& \teqb			& F		\\

\sidehead{\em Planetary parameters: \planetc}
Mass \mpl\ (\mearth)				& \mplanetc		& B,C,D	\\
Radius \rpl\ (\rearth)				& \rplanetc		& A,B,C,D	\\
Density \rhopl\ (\gcmc)				& \rhoplanetc		& A,B,C,D	\\
Orbital semi-major axis $a$ (AU)			& \semiMajc		& E		\\
Incident Flux $F$ (\fluxunit)			& $4.03 \times 10^8$    &A,C \\
Equilibrium temperature \teq\ (K)		& \teqc			& F		\\

\sidehead{\em Planetary parameters: \planetd}
Mass \mpl\ (\mearth)				& \mplanetd		& B,C,D	\\
Radius \rpl\ (\rearth)				& \rplanetd		& A,B,C,D	\\
Density \rhopl\ (\gcmc)				& \rhoplanetd		& A,B,C,D	\\
Orbital semi-major axis $a$ (AU)			& \semiMajd		& E		\\
Incident Flux $F$ (\fluxunit)			& $1.46 \times 10^8$    &A,C \\
Equilibrium temperature \teq\ (K)		& \teqd			& F		\\
\sidehead{\em Planetary parameters: \planete}
Mass \mpl\ (\mearth)				& \mplanete		& B,C,D	\\
Radius \rpl\ (\rearth)				& \rplanete		& A,B,C,D	\\
Density \rhopl\ (\gcmc)				& \rhoplanete		& A,B,C,D	\\
Orbital semi-major axis $a$ (AU)			& \semiMaje		& E		\\
Incident Flux $F$ (\fluxunit)			& $4.46 \times 10^8$    &A,C \\
Equilibrium temperature \teq\ (K)		& \teqe			& F		\\
\enddata
\tablecomments{
A: Based primarily on an analysis of the photometry,\\
B: Based on a joint analysis of the photometry and radial velocities,\\
C: Based on an analysis by D. Fischer of the Keck/HIRES template spectrum using SME \citep{Valenti1996},\\
D: Based on the Yale-Yonsei isochrones \citep{Yi2001} and the results from A, B, and C,\\
E: Based on Newton's revised version of Kepler's Third Law and the results from D,\\
F: Calculated assuming Bond albedos of 0.3 (b; Earth), 0.4 (c; Neptune), 0.34 (d; Jupiter), and 0.4 (e; Neptune) and complete redistribution of heat for reradiation.\\
}
\end{deluxetable*}

\begin{deluxetable*}{lrrrrll}
\tabletypesize{\tiny}
\tablewidth{0pt} 
\tablecaption{Exoplanets with Measured Mass and Radius \label{tab:all_planets}}
\tablehead{\colhead{Name} & \colhead{\mpl} & \colhead{\rpl} & \colhead{Incident Flux} & \colhead{Period} & \colhead{First Ref.} & \colhead{Orbit Ref.} \\ 
\colhead{} & \colhead{(\mearth)} & \colhead{(\rearth)} & \colhead{(\fluxunit)} & \colhead{(days)} & \colhead{} & \colhead{} } 

\startdata
            55 Cnc e &      7.862 &      2.078 &   3.34E+09 &      0.737 &                     \citet{McArthur2004} &                       \citet{Demory2011} \\ 
           CoRoT-1 b &    327.284 &     16.685 &   2.96E+09 &      1.509 &                        \citet{Barge2008} &                        \citet{Barge2008} \\ 
          CoRoT-10 b &    875.699 &     10.862 &   5.39E+07 &     13.241 &                       \citet{Bonomo2010} &                       \citet{Bonomo2010} \\ 
          CoRoT-11 b &    746.274 &     16.013 &   2.04E+09 &      2.994 &                     \citet{Gandolfi2010} &                     \citet{Gandolfi2010} \\ 
          CoRoT-12 b &    292.131 &     16.125 &   9.86E+08 &      2.828 &                       \citet{Gillon2010} &                       \citet{Gillon2010} \\ 
          CoRoT-13 b &    416.654 &      9.910 &   5.99E+08 &      4.035 &                      \citet{Cabrera2010} &                      \citet{Cabrera2010} \\ 
          CoRoT-14 b &   2445.485 &     12.205 &   3.29E+09 &      1.512 &                      \citet{Tingley2011} &                      \citet{Tingley2011} \\ 
          CoRoT-17 b &    781.832 &     11.422 &   1.32E+09 &      3.768 &                    \citet{Csizmadia2011} &                    \citet{Csizmadia2011} \\ 
          CoRoT-18 b &   1108.075 &     14.669 &   1.23E+09 &      1.900 &                      \citet{Hebrard2011} &                      \citet{Hebrard2011} \\ 
          CoRoT-19 b &    352.115 &     14.445 &   1.72E+09 &      3.897 &                     \citet{Guenther2012} &                     \citet{Guenther2012} \\ 
           CoRoT-2 b &   1041.197 &     16.405 &   1.27E+09 &      1.743 &                       \citet{Alonso2008} &                       \citet{Alonso2008} \\ 
           CoRoT-4 b &    228.008 &     13.325 &   3.02E+08 &      9.202 &           \citet{Moutou2008,Aigrain2008} &                       \citet{Moutou2008} \\ 
           CoRoT-5 b &    147.002 &     15.542 &   9.74E+08 &      4.038 &                        \citet{Rauer2009} &                        \citet{Rauer2009} \\ 
           CoRoT-6 b &    938.715 &     13.057 &   2.43E+08 &      8.887 &                     \citet{Fridlund2010} &                     \citet{Fridlund2010} \\ 
           CoRoT-7 b &      5.021 &      1.677 &   2.43E+09 &      0.854 &             \citet{Queloz2009,Leger2009} &                       \citet{Queloz2009} \\ 
           CoRoT-8 b &     68.673 &      6.383 &   1.21E+08 &      6.212 &                        \citet{Borde2010} &                        \citet{Borde2010} \\ 
           CoRoT-9 b &    268.099 &     11.758 &   6.59E+06 &     95.274 &                         \citet{Deeg2010} &                         \citet{Deeg2010} \\ 
           GJ 1214 b &      6.468 &      2.675 &   2.23E+07 &      1.580 &                  \citet{Charbonneau2009} &                       \citet{Carter2011} \\ 
            GJ 436 b &     23.105 &      4.218 &   4.09E+07 &      2.644 &                       \citet{Butler2004} &                       \citet{Maness2007} \\ 
           HAT-P-1 b &    169.196 &     13.908 &   6.58E+08 &      4.465 &                        \citet{Bakos2007a} &                        \citet{Bakos2007a} \\ 
          HAT-P-11 b &     26.231 &      4.725 &   1.33E+08 &      4.888 &                        \citet{Bakos2010} &                        \citet{Bakos2010} \\ 
          HAT-P-12 b &     66.997 &     10.739 &   1.91E+08 &      3.213 &                      \citet{Hartman2009} &                      \citet{Hartman2009} \\ 
          HAT-P-13 b &    272.394 &     14.344 &   1.67E+09 &      2.916 &                        \citet{Bakos2009a} &                         \citet{Winn2010} \\ 
          HAT-P-14 b &    710.648 &     12.877 &   1.37E+09 &      4.628 &                       \citet{Torres2010} &                       \citet{Torres2010} \\ 
          HAT-P-15 b &    620.231 &     12.004 &   1.51E+08 &     10.864 &                       \citet{Kovacs2010} &                       \citet{Kovacs2010} \\ 
          HAT-P-16 b &   1335.623 &     14.434 &   1.58E+09 &      2.776 &                     \citet{Buchhave2010} &                     \citet{Buchhave2010} \\ 
          HAT-P-17 b &    168.493 &     11.310 &   8.91E+07 &     10.339 &                       \citet{Howard2012} &                       \citet{Howard2012} \\ 
          HAT-P-18 b &     62.675 &     11.142 &   1.17E+08 &      5.508 &                      \citet{Hartman2011a} &                      \citet{Hartman2011a} \\ 
          HAT-P-19 b &     92.889 &     12.676 &   2.36E+08 &      4.009 &                      \citet{Hartman2011a} &                      \citet{Hartman2011a} \\ 
           HAT-P-2 b &   2819.241 &     12.956 &   1.10E+09 &      5.633 &                        \citet{Bakos2007b} &                          \citet{Pal2010} \\ 
          HAT-P-20 b &   2316.734 &      9.708 &   2.02E+08 &      2.875 &                        \citet{Bakos2011} &                        \citet{Bakos2011} \\ 
          HAT-P-21 b &   1296.160 &     11.466 &   6.12E+08 &      4.124 &                        \citet{Bakos2011} &                        \citet{Bakos2011} \\ 
          HAT-P-22 b &    683.741 &     12.094 &   6.12E+08 &      3.212 &                        \citet{Bakos2011} &                        \citet{Bakos2011} \\ 
          HAT-P-23 b &    666.163 &     15.318 &   4.03E+09 &      1.213 &                        \citet{Bakos2011} &                        \citet{Bakos2011} \\ 
          HAT-P-24 b &    217.977 &     13.908 &   1.63E+09 &      3.355 &                      \citet{Kipping2010} &                      \citet{Kipping2010} \\ 
          HAT-P-26 b &     18.640 &      6.327 &   2.23E+08 &      4.235 &                      \citet{Hartman2011b} &                      \citet{Hartman2011b} \\ 
          HAT-P-27 b &    195.955 &     11.623 &   4.34E+08 &      3.040 &                     \citet{Anderson2011a} &                     \citet{Anderson2011a} \\ 
          HAT-P-28 b &    199.536 &     13.572 &   8.27E+08 &      3.257 &                     \citet{Buchhave2011a} &                     \citet{Buchhave2011a} \\ 
          HAT-P-29 b &    247.580 &     12.396 &   5.70E+08 &      5.723 &                     \citet{Buchhave2011a} &                     \citet{Buchhave2011a} \\ 
           HAT-P-3 b &    189.327 &     10.067 &   4.08E+08 &      2.900 &                       \citet{Torres2007} &                       \citet{Torres2008} \\ 
          HAT-P-30 b &    225.996 &     15.005 &   1.63E+09 &      2.811 &                      \citet{Johnson2011} &                      \citet{Johnson2011} \\ 
          HAT-P-31 b &    689.358 &     11.982 &   8.46E+08 &      5.005 &                      \citet{Kipping2011} &                      \citet{Kipping2011} \\ 
          HAT-P-32 b &    302.182 &     22.810 &   2.62E+09 &      2.150 &                      \citet{Hartman2011c} &                      \citet{Hartman2011c} \\ 
          HAT-P-33 b &    243.556 &     20.458 &   2.60E+09 &      3.474 &                      \citet{Hartman2011c} &                      \citet{Hartman2011c} \\ 
          HAT-P-34 b &   1059.600 &     13.404 &   7.35E+08 &      5.453 &                        \citet{Bakos2012} &                        \citet{Bakos2012} \\ 
          HAT-P-35 b &    335.047 &     14.915 &   1.41E+09 &      3.647 &                        \citet{Bakos2012} &                        \citet{Bakos2012} \\ 
          HAT-P-36 b &    584.539 &     14.154 &   2.49E+09 &      1.327 &                        \citet{Bakos2012} &                        \citet{Bakos2012} \\ 
          HAT-P-37 b &    372.996 &     13.191 &   6.01E+08 &      2.797 &                        \citet{Bakos2012} &                        \citet{Bakos2012} \\ 
           HAT-P-4 b &    213.416 &     14.266 &   1.87E+09 &      3.057 &                       \citet{Kovacs2007} &                       \citet{Kovacs2007} \\ 
           HAT-P-5 b &    335.490 &     14.042 &   1.27E+09 &      2.788 &                        \citet{Bakos2007c} &                        \citet{Bakos2007c} \\ 
           HAT-P-6 b &    336.749 &     14.893 &   1.78E+09 &      3.853 &                        \citet{Noyes2008} &                        \citet{Noyes2008} \\ 
           HAT-P-7 b &    572.656 &     15.262 &   5.57E+09 &      2.205 &                          \citet{Pal2008} &                         \citet{Winn2009} \\ 
           HAT-P-8 b &    411.052 &     16.797 &   2.24E+09 &      3.076 &                       \citet{Latham2009} &                       \citet{Latham2009} \\ 
           HAT-P-9 b &    246.817 &     15.677 &   1.24E+09 &      3.923 &                      \citet{Shporer2009} &                      \citet{Shporer2009} \\ 
         HD 149026 b &    114.882 &      7.323 &   1.78E+09 &      2.876 &                         \citet{Sato2005} &                       \citet{Carter2009} \\ 
          HD 17156 b &   1049.670 &     11.422 &   1.95E+08 &     21.217 &                      \citet{Fischer2007} &                     \citet{Barbieri2009} \\ 
         HD 189733 b &    363.454 &     12.743 &   4.71E+08 &      2.219 &                       \citet{Bouchy2005} &                       \citet{Bouchy2005} \\ 
         HD 209458 b &    219.181 &     15.218 &   9.93E+08 &      3.525 &        \citet{Henry2000,Charbonneau2000} &                       \citet{Torres2008} \\ 
          HD 80606 b &   1236.479 &     11.522 &   1.59E+07 &    111.437 &                         \citet{Naef2001} &                       \citet{Moutou2009} \\ 
           KOI-135 b &   1027.001 &     13.437 &   1.63E+09 &      3.024 &                      \citet{Borucki2011} &                       \citet{Bonomo2012} \\ 
           KOI-196 b &    156.857 &      9.417 &   1.40E+09 &      1.856 &                      \citet{Borucki2011} &                     \citet{Santerne2011} \\ 
           KOI-204 b &    324.519 &     13.885 &   1.51E+09 &      3.247 &                      \citet{Borucki2011} &                       \citet{Bonomo2012} \\ 
           KOI-254 b &    162.563 &     10.750 &   8.91E+07 &      2.455 &                      \citet{Borucki2011} &                      \citet{Johnson2012} \\ 
           KOI-428 b &    691.836 &     13.101 &   1.54E+09 &      6.873 &                     \citet{Santerne2011} &                     \citet{Santerne2011} \\ 
         Kepler-10 b &      4.539 &      1.415 &   4.88E+09 &      0.837 &                      \citet{Batalha2011} &                      \citet{Batalha2011} \\ 
         Kepler-11 b &      4.298 &      1.968 &   1.86E+08 &     10.304 &                     \citet{Lissauer2011} &                     \citet{Lissauer2011} \\ 
         Kepler-11 c &     13.500 &      3.147 &   1.36E+08 &     13.025 &                     \citet{Lissauer2011} &                     \citet{Lissauer2011} \\ 
         Kepler-11 d &      6.100 &      3.427 &   6.50E+07 &     22.687 &                     \citet{Lissauer2011} &                     \citet{Lissauer2011} \\ 
         Kepler-11 e &      8.401 &      4.515 &   4.11E+07 &     31.996 &                     \citet{Lissauer2011} &                     \citet{Lissauer2011} \\ 
         Kepler-11 f &      2.298 &      2.607 &   2.48E+07 &     46.689 &                     \citet{Lissauer2011} &                     \citet{Lissauer2011} \\ 
         Kepler-12 b &    137.283 &     18.980 &   1.09E+09 &      4.438 &                      \citet{Borucki2011} &                      \citet{Fortney2011} \\ 
         Kepler-14 b &   2671.703 &     12.721 &   1.32E+09 &      6.790 &                      \citet{Borucki2011} &                     \citet{Buchhave2011b} \\ 
         Kepler-15 b &    210.532 &     10.750 &   3.45E+08 &      4.943 &                      \citet{Borucki2011} &                         \citet{Endl2011} \\ 
         Kepler-16 b &    105.833 &      8.441 &   4.84E+05 &    228.776 &                      \citet{Borucki2011} &                        \citet{Doyle2011} \\ 
         Kepler-17 b &    788.004 &     14.893 &   2.10E+09 &      1.486 &                      \citet{Borucki2011} &                       \citet{Desert2011} \\ 
         Kepler-18 b &      6.900 &      5.484 &   6.32E+08 &      3.505 &                      \citet{Borucki2011} &                      \citet{Cochran2011} \\ 
         Kepler-18 c &     17.299 &      5.484 &   2.24E+08 &      7.642 &                      \citet{Borucki2011} &                      \citet{Cochran2011} \\ 
         Kepler-18 d &     16.399 &      6.973 &   9.21E+07 &     14.859 &                      \citet{Borucki2011} &                      \citet{Cochran2011} \\ 
         Kepler-20 b &      8.474 &      1.906 &   4.70E+08 &      3.696 &                      \citet{Borucki2011} &                      \citet{Gautier2012} \\ 
         Kepler-20 c &     15.734 &      3.064 &   1.12E+08 &     10.854 &                      \citet{Borucki2011} &                      \citet{Gautier2012} \\ 
         Kepler-20 d &      7.528 &      2.745 &   8.12E+06 &     77.612 &                      \citet{Borucki2011} &                      \citet{Gautier2012} \\ 
          Kepler-4 b &     24.544 &      3.998 &   1.54E+09 &      3.213 &                      \citet{Borucki2010} &                      \citet{Borucki2010} \\ 
          Kepler-5 b &    672.699 &     16.024 &   2.42E+09 &      3.548 &                         \citet{Koch2010} &                         \citet{Koch2010} \\ 
          Kepler-6 b &    212.739 &     14.815 &   1.16E+09 &      3.235 &                       \citet{Dunham2010} &                       \citet{Dunham2010} \\ 
          Kepler-7 b &    139.127 &     16.550 &   1.33E+09 &      4.886 &                       \citet{Latham2010} &                       \citet{Latham2010} \\ 
          Kepler-8 b &    186.158 &     15.890 &   1.73E+09 &      3.523 &                      \citet{Jenkins2010} &                      \citet{Jenkins2010} \\ 
       OGLE-TR-182 b &    325.603 &     12.653 &   7.45E+08 &      3.979 &                         \citet{Pont2008} &                         \citet{Pont2008} \\ 
       OGLE-TR-211 b &    240.675 &     15.229 &   2.01E+09 &      3.677 &                      \citet{Udalski2008} &                      \citet{Udalski2008} \\ 
       OGLE2-TR-L9 b &   1453.828 &     18.028 &   3.89E+09 &      2.486 &                      \citet{Snellen2009} &                      \citet{Snellen2009} \\ 
           Qatar-1 b &    346.352 &     13.034 &   8.45E+08 &      1.420 &                      \citet{Alsubai2011} &                      \citet{Alsubai2011} \\ 
            TrES-1 b &    239.152 &     11.948 &   3.88E+08 &      3.030 &                       \citet{Alonso2004} &                       \citet{Alonso2004} \\ 
            TrES-2 b &    381.607 &     13.706 &   1.14E+09 &      2.471 &                    \citet{O'Donovan2006} &                    \citet{O'Donovan2006} \\ 
            TrES-4 b &    292.473 &     19.607 &   2.31E+09 &      3.554 &                    \citet{Mandushev2007} &                    \citet{Mandushev2007} \\ 
            TrES-5 b &    565.109 &     13.538 &   1.09E+09 &      1.482 &                    \citet{Mandushev2011} &                    \citet{Mandushev2011} \\ 
            WASP-1 b &    263.078 &     16.976 &   2.65E+09 &      2.520 &              \citet{CollierCameron2007} &                      \citet{Simpson2011b} \\ 
           WASP-10 b &   1013.770 &     12.094 &   2.38E+08 &      3.093 &                    \citet{Christian2009} &                    \citet{Christian2009} \\ 
           WASP-11 b &    171.543 &     10.190 &   1.87E+08 &      3.722 &               \citet{West2009a,Bakos2009b} &                         \citet{West2009a} \\ 
           WASP-12 b &    432.432 &     20.044 &   1.01E+10 &      1.091 &                         \citet{Hebb2009} &                  \citet{Maciejewski2011} \\ 
           WASP-13 b &    152.357 &     15.554 &   1.12E+09 &      4.353 &                      \citet{Skillen2009} &                      \citet{Skillen2009} \\ 
           WASP-14 b &   2444.754 &     14.344 &   2.75E+09 &      2.244 &                        \citet{Joshi2009} &                        \citet{Joshi2009} \\ 
           WASP-15 b &    172.613 &     15.990 &   1.69E+09 &      3.752 &                         \citet{West2009b} &                         \citet{West2009b} \\ 
           WASP-16 b &    267.695 &     11.287 &   6.64E+08 &      3.119 &                       \citet{Lister2009} &                       \citet{Lister2009} \\ 
           WASP-17 b &    156.828 &     16.909 &   1.34E+09 &      3.735 &                     \citet{Anderson2010} &                     \citet{Anderson2010} \\ 
           WASP-18 b &   3206.179 &     12.385 &   7.50E+09 &      0.941 &                      \citet{Hellier2009a} &                      \citet{Hellier2009a} \\ 
           WASP-19 b &    360.211 &     15.520 &   4.13E+09 &      0.789 &                         \citet{Hebb2010} &                      \citet{Hellier2011} \\ 
            WASP-2 b &    288.782 &     11.993 &   6.47E+08 &      2.152 &              \citet{CollierCameron2007} &                  \citet{Charbonneau2007} \\ 
           WASP-21 b &     95.431 &     11.982 &   5.75E+08 &      4.322 &                       \citet{Bouchy2010} &                       \citet{Bouchy2010} \\ 
           WASP-22 b &    177.678 &     12.541 &   9.25E+08 &      3.533 &                       \citet{Maxted2010a} &                       \citet{Maxted2010a} \\ 
           WASP-23 b &    277.208 &     10.772 &   3.69E+08 &      2.944 &                       \citet{Triaud2011} &                       \citet{Triaud2011} \\ 
           WASP-24 b &    346.738 &     14.557 &   2.22E+09 &      2.341 &                       \citet{Street2010} &                      \citet{Simpson2011b} \\ 
           WASP-25 b &    183.847 &     13.661 &   5.06E+08 &      3.765 &                        \citet{Enoch2011} &                        \citet{Enoch2011} \\ 
           WASP-26 b &    323.366 &     14.781 &   9.03E+08 &      2.757 &                      \citet{Smalley2010} &                      \citet{Smalley2010} \\ 
           WASP-29 b &     77.261 &      8.869 &   2.04E+08 &      3.923 &                      \citet{Hellier2010} &                      \citet{Hellier2010} \\ 
            WASP-3 b &    639.396 &     14.445 &   3.56E+09 &      1.847 &                     \citet{Pollacco2008} &                     \citet{Tripathi2010} \\ 
           WASP-31 b &    152.339 &     17.211 &   1.40E+09 &      3.406 &                     \citet{Anderson2011b} &                     \citet{Anderson2011b} \\ 
           WASP-32 b &   1129.221 &     13.213 &   1.35E+09 &      2.719 &                       \citet{Maxted2010b} &                       \citet{Maxted2010b} \\ 
           WASP-34 b &    185.399 &     13.661 &   4.13E+08 &      4.318 &                      \citet{Smalley2011} &                      \citet{Smalley2011} \\ 
           WASP-35 b &    227.906 &     14.781 &   1.01E+09 &      3.162 &                        \citet{Enoch2011} &                        \citet{Enoch2011} \\ 
           WASP-36 b &    721.220 &     14.210 &   1.89E+09 &      1.537 &                        \citet{Smith2012} &                        \citet{Smith2012} \\ 
           WASP-37 b &    570.041 &     12.989 &   7.01E+08 &      3.577 &                      \citet{Simpson2011a} &                      \citet{Simpson2011a} \\ 
           WASP-38 b &    854.794 &     12.250 &   5.49E+08 &      6.872 &                       \citet{Barros2011} &                       \citet{Barros2011} \\ 
           WASP-39 b &     90.367 &     14.221 &   3.54E+08 &      4.055 &                        \citet{Faedi2011} &                        \citet{Faedi2011} \\ 
            WASP-4 b &    388.827 &     15.352 &   1.77E+09 &      1.338 &                       \citet{Wilson2008} &                       \citet{Wilson2008} \\ 
           WASP-41 b &    296.362 &     13.437 &   5.35E+08 &      3.052 &                       \citet{Maxted2011} &                       \citet{Maxted2011} \\ 
           WASP-43 b &    564.470 &     10.414 &   8.11E+08 &      0.813 &                      \citet{Hellier2011} &                      \citet{Hellier2011} \\ 
           WASP-48 b &    312.807 &     18.700 &   3.88E+09 &      2.144 &                        \citet{Enoch2011} &                        \citet{Enoch2011} \\ 
            WASP-5 b &    516.398 &     13.112 &   2.09E+09 &      1.628 &                     \citet{Anderson2008} &                     \citet{Anderson2008} \\ 
           WASP-50 b &    467.967 &     12.911 &   8.54E+08 &      1.955 &                       \citet{Gillon2011} &                       \citet{Gillon2011} \\ 
            WASP-6 b &    165.697 &     13.706 &   4.46E+08 &      3.361 &                       \citet{Gillon2009} &                       \citet{Gillon2009} \\ 
            WASP-7 b &    292.162 &     10.246 &   8.60E+08 &      4.955 &                      \citet{Hellier2009b} &                      \citet{Hellier2009b} \\ 
            WASP-8 b &    679.482 &     11.623 &   1.76E+08 &      8.159 &                       \citet{Queloz2010} &                       \citet{Queloz2010} \\ 
              XO-1 b &    291.903 &     13.504 &   4.82E+08 &      3.942 &                   \citet{McCullough2006} &                   \citet{McCullough2006} \\ 
              XO-2 b &    180.056 &     11.007 &   6.93E+08 &      2.616 &                        \citet{Burke2007} &                        \citet{Burke2007} \\ 
              XO-5 b &    366.286 &     11.534 &   4.82E+08 &      4.188 &                        \citet{Burke2008} &                        \citet{Burke2008} \\ 
            KOI-94 b &      9.400 &      1.770 &   1.58E+09 &      3.743 &                      \citet{Borucki2011} &                        Weiss et al. (2013) \\ 
            KOI-94 c &      8.300 &      4.280 &   4.03E+08 &     10.424 &                      \citet{Borucki2011} &                        Weiss et al. (2013) \\ 
            KOI-94 d &    105.000 &     11.400 &   1.46E+08 &     22.343 &                      \citet{Borucki2011} &                       Weiss et al. (2013) \\ 
            KOI-94 e &     38.000 &      6.640 &   4.46E+07 &     54.320 &                        \citet{Batalha2012} &                        Weiss et al. (2013)\\ 
\hline
               Earth &      1.000 &      1.000 &   1.07E+06 &    365.250 &                               &                              \\ 
             Jupiter &    317.817 &     11.198 &   3.97E+04 &   4336.069 &                               &                              \\ 
              Saturn &     95.027 &      9.440 &   1.17E+04 &  10833.641 &                               &                              \\ 
              Uranus &     14.535 &      4.003 &   2.91E+03 &  30730.951 &                               &                             \\ 
             Neptune &     17.147 &      3.879 &   1.19E+03 &  60157.796 &                               &                             \\ 
\enddata
\end{deluxetable*}
\end{document}